\documentclass[twocolumn,showpacs,pre,superscriptaddress,floatfix]{revtex4}

\usepackage{graphicx}
\usepackage{amsmath}
\usepackage{amssymb}
\usepackage{bm}
\usepackage{times}
\usepackage[pdftex]{hyperref}
\hypersetup{colorlinks=true,linkcolor=blue,citecolor=blue,urlcolor=blue}

\newcommand{\Df}{\Delta F}
\newcommand{\DfPAP}{\Delta F_{\rm{P,AP}}}
\newcommand{\Hrep}{{\cal H}_{\mathrm{rep}}}
\newcommand{\ab}{_{\alpha\beta}}
 
\newcommand{\pbc}{\pi}
\newcommand{\abc}{\overline{\pi}}

\begin{document}

\title{Interface free-energy exponent in the one-dimensional  Ising
spin  glass  with long-range  interactions  in both  the  droplet  and
broken  replica symmetry   regions}

\author{T.~Aspelmeier}
\affiliation{Felix Bernstein Institute for Mathematical Statistics in
the Biosciences, Georg August University of G\"{o}ttingen, 37077
G\"{o}ttingen, Germany}
\affiliation{Institute for Mathematical Stochastics, University of
G\"{o}ttingen, 37073 G\"{o}ttingen, Germany}
\affiliation{Statistical Inverse Problems in Biophysics, Max Planck
Institute for Biophysical Chemistry, 37077 G\"{o}ttingen, Germany}

\author{Wenlong Wang}
\affiliation{Department of Physics and Astronomy, Texas A\&M University,
College Station, Texas 77843-4242, USA}
\date{\today}

\author{M.~A.~Moore}
\affiliation{School of Physics and Astronomy, University of Manchester,
Manchester M13 9PL, United Kingdom}

\author{Helmut G.~Katzgraber}
\affiliation{Department of Physics and Astronomy, Texas A\&M University,
College Station, Texas 77843-4242, USA}
\affiliation{Santa Fe Institute, 1399 Hyde Park Road, Santa Fe, New
Mexico 87501, USA}

\begin{abstract}

The one-dimensional  Ising spin-glass model  with power-law long-range
interactions is a useful proxy model for studying  spin glasses in
higher space dimensions  and for finding the dimension at which the
spin-glass state changes from having broken replica  symmetry to that of
droplet  behavior. To  this end  we have calculated  the exponent that
describes the  difference in free energy between periodic  and
antiperiodic  boundary conditions.  Numerical  work is  done to support
some of the assumptions made  in the calculations and to determine the
behavior of the interface free-energy exponent of the power law of the
interactions.  Our numerical results for the interface free-energy
exponent are badly affected by finite-size problems.

\end{abstract}

\maketitle

The  Edwards-Anderson  (EA)   Hamiltonian  \cite{edwards:75} is
universally agreed  to capture the essence of  spin-glass behavior.
However,  what is not  agreed upon is the nature  of its low-temperature
ordered state. There  are two main theories. The first is the replica
symmetry breaking  (RSB) theory of Parisi
\cite{parisi:79,parisi:80,parisi:83,rammal:86,mezard:87,parisi:08},
which is known  to be  correct for  the  Sherrington-Kirkpatrick (SK)
model \cite{sherrington:75},  which  is  the mean-field  or
infinite-dimensional limit of the EA model. It is characterized by a
very large number of  pure states that organize into an ultrametric
topology \cite{mezard:87}. On the  other hand, in the  second theory,
the droplet  picture, developed in
Refs.~\cite{mcmillan:84,bray:86,fisher:86,fisher:87,fisher:88}, there
are only  two pure states. In  this  picture  behavior is dominated by
low-lying excitations or droplets whose (free) energies scale as their
linear  dimension $\ell$ as $\ell^{\theta}$  and have  a fractal
dimension $d_s$ where $d-1  < d_s <d$ for a $d$-dimensional system. In
contrast, in  the RSB  picture there  are low-lying excitations that
cost  an energy  of $O(1)$  and are space  filling, that is, $d_s=d$.
Despite the  striking differences of the two pictures, it has proven
difficult to establish by either experiment or simulations which holds
for, say, three-dimensional ($d=3$) spin glasses.

Much of the effort in this regard has focused on the existence or
absence of  the  de Almeida Thouless (AT)  line \cite{almeida:78} that
separates a spin-glass state in a field from a paramagnetic state.   In
the  RSB picture for Ising  spin glasses (only these will be discussed
in this paper), there is a phase transition  in the field  $h$ and
temperature $T$ plane separating the paramagnetic  phase from a phase
with RSB. In the  droplet  picture, the application   of  a  field
removes the  phase transition to  the spin-glass phase, which then
occurs only in zero field, just   as   for the  Ising   ferromagnet.
We   have argued \cite{moore:11} that  there is an AT line for
dimensions $d  > 6$ and that for $d  \le 6$ the droplet picture applies
and  the AT line is absent. The calculation involved determining the
form of this line in the limit as $T \rightarrow  T_c$ but  what one
really needs is  to show  that for \textit{any} $T <T_c$, there is no
transition in a field. An attempt was made to do this using a $1/m$
expansion for an $m$-component random field added to the $m$-component
EA vector model \cite{moore:12}, and once again $d=6$ emerged as the
dimension below which  the droplet picture might be appropriate, but the
argument is rather convoluted. A tentative argument that  there might be
no  AT line when  $d \le 6$ was  made by Bray and  Roberts
\cite{bray:80b}  when they were  unable to  find any stable
perturbative fixed  points  in an $\epsilon$-expansion where
$d=6-\epsilon$. Suggestive though these arguments, which are based on the
form  of the AT line or  the critical exponents across it, are, they do
not get really to the heart of the matter, which is the nature of the
low-temperature phase in spin glasses. This is controlled by a
zero-temperature fixed  point, rather than  a critical fixed point. In
this paper we focus on this zero-temperature fixed point and its
associated exponent $\theta$.

While  we believe  that $d=6$  is the  dimension below  which  the
low-temperature phase  is as  described by the  droplet picture  and
above which for $d > 6$ by RSB  ideas, there  is clearly  little chance
that numerical studies  could  be  done  in  such high  space dimensions
to  confirm  this changeover. However, it  is possible  to imagine
numerical  work to confirm the  equivalent changeover  in the
one-dimensional  Ising spin-glass   model  introduced   by  Kotliar
{\em et al.}~\cite{kotliar:83} given by the Hamiltonian
\begin{equation}
\mathcal{H}=-\sum_{i<j} J_{ij} S_i S_j,
\label{eqn:KAS}
\end{equation}
where the Ising spins $S_i=\pm 1$ are distributed on a one-dimensional
ring of length $L$ to enforce periodic boundary conditions. The
interactions $J_{ij}$ are specified by
\begin{equation}
J_{ij}=c(\sigma) \frac{\epsilon_{ij}}{r_{ij}^{\sigma}},
\label{eqn:Jij}
\end{equation}
where \cite{katzgraber:03}
\begin{equation}
r_{ij}= \frac{L}{\pi}\sin\left(\frac{ \pi|i-j|}{L}\right)
\label{eqn:distance}
\end{equation}
is the chord between sites $i$ and $j$. The disorder $\epsilon_{ij}$ is
chosen according to a Gaussian distribution of zero mean and standard
deviation unity, while the constant $c(\sigma)$ in Eq.~(\ref{eqn:Jij})
is fixed to make the mean-field transition temperature
$T_c^{\mathrm{MF}}=1$, where  $[\cdots]_{\rm av}$ represents a disorder
average so that $[J_{ij}^2]_{\rm av}=c(\sigma)^2/r_{ij}^{2 \sigma}$.
Here $(T_c^{\mathrm{MF}})^2=  \sum_j [J_{ij}^2]_{\mathrm{av}}$. We will take
$[J_{ii}^2]_{\mathrm{av}}=0$.

The  phase diagram  of this  model in the $d$--$\sigma$  plane  has been
deduced      from     renormalization      group      arguments     in
Refs.~\cite{fisher:88,bray:86b,katzgraber:03}. For  $d   =1$, the
model behaves just like the SK model  when  $0 \le \sigma <1/2$.  For
$1/2< \sigma <2/3$ the critical exponents at the spin-glass transition
are mean-field like, but in the interval $2/3 \le \sigma <1$, the
critical exponents are changed by fluctuations away from their
mean-field values. When $\sigma \ge 1$, $T_c(\sigma)=0$. There is a
convenient mapping between $\sigma$ and an effective dimensionality
$d_{\textrm{eff}}$ of the short-range EA model
\cite{katzgraber:03,katzgraber:09b,leuzzi:09,banos:12b,aspelmeier:16}.
For $1/2 < \sigma < 2/3$, it is
\begin{equation}
d_{\textrm{eff}}=\frac{2}{2 \sigma-1}.
\label{eqn:def}
\end{equation}
Thus, right at the value of $\sigma =2/3$, $d_{\textrm{eff}}=6$. This
mapping has a precise sense  for equations associated with finite-size
critical scaling at least when  $1/2 < \sigma  < 2/3$. Whereas  for the
short-range EA  model   there   is  an   expression  involving   the
dimensionality  $d$, the corresponding  formula for  the
Kotliar-Anderson-Stein (KAS)  model is
obtained   by    replacing   $d$   by    the effective space dimension
$d_{\textrm{eff}}$   of Eq.~(\ref{eqn:def}) \cite{aspelmeier:16}.

In Ref.~\cite{moore:11} it was  shown that the  arguments that 
had led us  to believe that $6$ is the  lower critical dimension for
replica symmetry breaking, such as the  form of the AT line near $T_c$
and the  Bray-Roberts study  of the critical  exponents across  the AT
line, suggested  also that  $\sigma=2/3$  was  the  special value  of
$\sigma$ for  the KAS  model. Thus, we  suspect that for  $\sigma <2/3$
there is RSB in the low-temperature  phase, while for  $1> \sigma \ge
2/3$  there is droplet  behavior. The  purpose  of this  paper is  to
strengthen  these  arguments   by  calculating  the  exponent $\theta$
of the zero-temperature fixed point. That we can do this  is another
advantage of the  KAS model. In  the droplet region,    it    has   been
realized    for    many   years    that $\theta=1-\sigma$
\cite{bray:86b,fisher:88,moore:10}.  We  will   argue  below that
$\theta=1/6$  in the  RSB region, if  one defines  $\theta$   from  the
variance   of  the sample-to-sample   free-energy   differences  between
periodic  and antiperiodic boundary conditions.   For the EA model
$\theta$ and $d_s$ in the  droplet regime are only known  from numerical
studies or simple renormalization-group  approximations
\cite{southern:77},  in particular that of Migdal and Kadanoff
\cite{bray:84}.

\begin{figure}[htb]
\begin{center}
\includegraphics[width=\columnwidth]{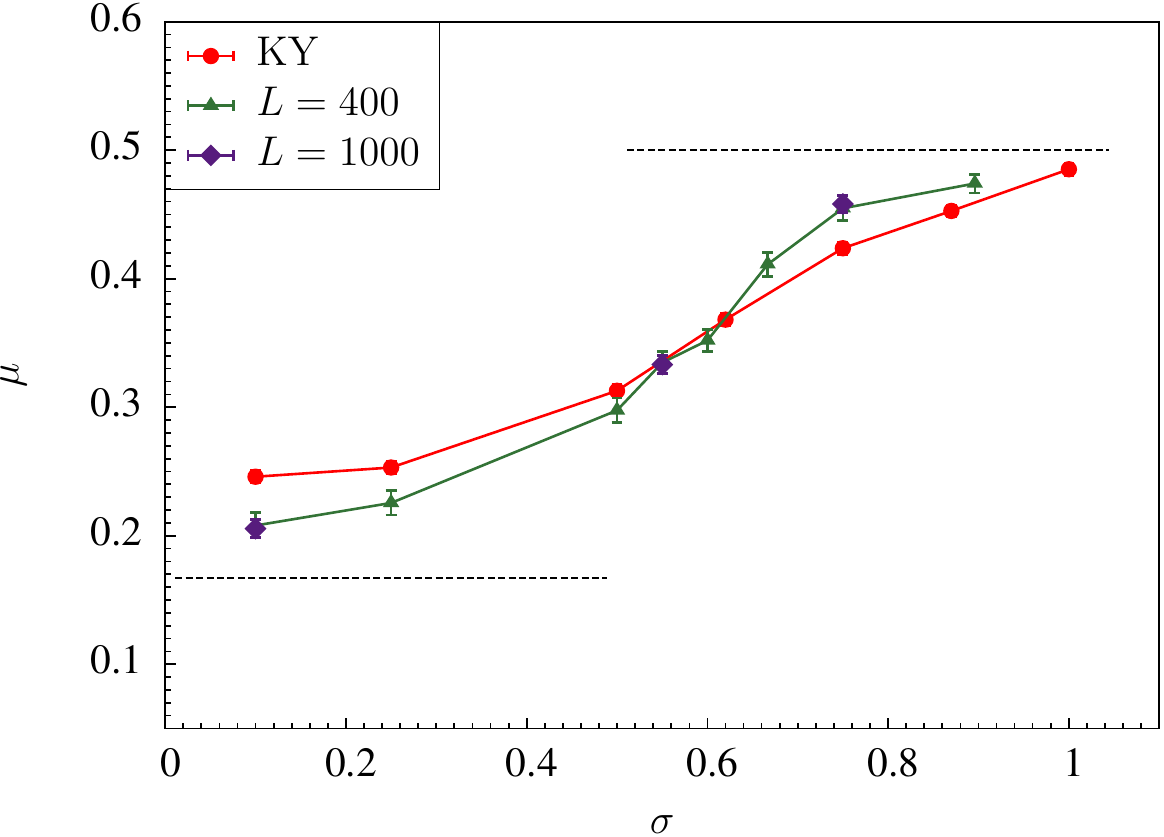}
\caption{(Color online) Estimates of the exponent $\mu$
(sample-to-sample variation of the ground-state energy) as a function of
$\sigma$, i.e., $\delta E \sim L^{\mu}$. Here KY denotes results obtained on
samples up to $L =256$ by Katzgraber and Young (KY)
\cite{katzgraber:03}.  The expectation for $\mu$ is that $\mu=1/6$ for
$\sigma <1/2$ \cite{parisi:10} and $1/2$ for all $\sigma > 1/2$, as
the system is now self-averaging \cite{wehr:90} (dashed lines).
Finite-size effects make the transition between these two values
for $\mu$ spread over a large range of $\sigma$. Notice that the results
for $\mu$ in the SK model region $\sigma =0.1$ are moving closer to
the theoretical prediction of $1/6$ as $L$ increases.}
\label{fig:mu}
\end{center}
\end{figure}

While, in  principle, the KAS model  allows one to do  numerical work on
systems that  might be the analog of high-dimensional hypercubic
systems of the EA model, there are problems with its use. Finite-size
effects are both large and difficult to understand and deal with.  To
illustrate this, we show  in Fig.~\ref{fig:mu} a plot of the exponent
$\mu$ (which describes the sample-to-sample variation, i.e., $\delta E
\sim L^{\mu}$), $\delta E$ of the  ground-state  energy  of  the  system
as a  function  of $\sigma$.  The estimate of $\mu$ is obtained by  just
fitting $\delta E$ to $L^{\mu}$, ignoring any corrections to scaling.
Clearly, the data are a long way from being satisfactorily fitted  by
this simple form, but if one is optimistic,  one could imagine that  as
$L$ is increased   the results tend  towards   the   theoretical
expectation. However, the improvement is  so slow   we   worried whether
the theoretical expectation that for the SK limit $\mu =1/6$
\cite{parisi:10} might  not be correct. In Appendix \ref{sec:proof} we
therefore have outlined a ``rigorous'' proof that  at least $\mu \le
1/5$. (We put rigorous in quotes to  indicate to that the proof cannot
be considered mathematically rigorous as  it involves  the  use of  the
replica trick.) In  this paper we need the value of  $\mu$ as we argue
that for all $\sigma < 2/3$, $\theta$ takes the SK limit value of $\mu$.

\section{The interface free energy}
\label{sec:theta}

One of the key concepts in the droplet picture of spin glasses is the
interface free energy
\cite{banavar:82,mcmillan:84,bray:86,fisher:86,fisher:88} $\delta F$
and the associated stiffness exponent $\theta$ defined by
\begin{equation}
\delta F \sim \ell^\theta ,
\end{equation}
where $\ell$ is the length scale of the excitation or droplet (or region
of flipped spins). If $\theta>0$, the spin-glass state is stable at
finite temperature, whereas if $\theta<0$, at $T=0$ large-scale
excitations cost little energy, so the spin-glass state is unstable at
finite temperature. Thus, the dimension or value of $\sigma$ at which
$\theta=0$  determines the lower critical dimension of the spin glass.
In this section we calculate $\theta$ \textit{analytically} first in the
RSB region $ \sigma < 2/3$ and then in the droplet region ($2/3 \le
\sigma < 1$) for the KAS model using the replica method and the
formalism of Ref.~\cite{aspelmeier:02}.

There are many ways of defining a droplet free-energy cost, but in this
section we will take it to be the interface free energy defined as the
root-mean-square change in the free energy of a spin glass when the
boundary conditions along one direction (the $z$ direction) are changed
from periodic to antiperiodic, i.e.,
\begin{equation}
\delta F=\sqrt{\overline{\DfPAP^2}} .
\end{equation}
Here and in the following, the overbar represents averaging over bond
configurations, where $\DfPAP = F_{\rm P} - F_{\rm AP}$, and $F_{\rm P}$
and $F_{\rm AP}$ are the free energies with periodic and antiperiodic
boundary conditions, respectively. Antiperiodic boundary conditions can
be realized by reversing the sign of the bonds crossing a diameter of
the ring in the KAS model. It follows that $\overline{\DfPAP}=
0$.

The basic strategy of Ref.~\cite{aspelmeier:02} was to replicate the
system with periodic boundary conditions $n$ times and the system with
antiperiodic boundary conditions $m$ times and keep $n$ distinct from
$m$. Expanding the replicated partition function in powers of $m$ and
$n$ and taking the logarithm, we obtain
\begin{multline}
-\ln \overline{ Z_{\rm P}^n Z_{\rm AP}^m } = 
(n + m) \, \beta \overline{F}  \\
- {(n+m)^2\over 2}\, \beta^2 \overline{\Df^2}
+ {nm \over 2} \, \beta^2 \overline{\DfPAP^2} +  \cdots ,
\label{lnZnZm}
\end{multline}
where $ \overline{\Df^2} = \overline{F_{\rm P}^2} - \overline{F_{\rm
P}}^2 = \overline{F_{\rm AP}^2} - \overline{F_{\rm AP}}^2 $ is the 
(mean-square) sample-to-sample fluctuation of the free energy, the same for
both sets of boundary conditions P or AP, and $\overline{F} =
\overline{F_{\rm P}} = \overline{F_{\rm AP}}$. Hence, to find the
variance of the interface free energy $\overline{\DfPAP^2}$ (which
scales with $L$ as $L^{2 \theta}$), we expand $\ln \overline{ Z_{\rm
P}^n Z_{\rm AP}^m }$ to second order in the numbers of replicas $n$ and
$m$, separate out the pieces involving the \textit{total} number of
replicas $n+m$, and take the remaining piece, which is proportional to
$n m$.

Using the standard replica field theory \cite{dedominicis:98}, we write
\begin{equation}
\overline{ Z_{\rm P}^n Z_{\rm AP}^m } = \int\mathcal{D}q\,\exp(-\beta\Hrep) , 
\end{equation}
where $\Hrep$ is the replica free energy, expressed in terms of the
spin-glass order parameter field $q\ab(x)$. For the short-range KAS
model it is given by
\begin{multline}
\beta \Hrep = \int dz \, \left[
-{\tau \over 2} \sum_{\alpha,\beta} q\ab^2 +
{1 \over 4} \sum_{\alpha,\beta} (\partial q\ab/\partial z)^2  \right. \\
\left.
-{w \over 6} \sum_{\alpha,\beta,\gamma}  q\ab q_{\beta\gamma}q_{\gamma\alpha} 
-{y \over 12} \sum_{\alpha,\beta} q\ab^4 
\right] ,
\label{Hrep}
\end{multline}
where $q_{\alpha\beta}$ is a symmetric matrix with $q_{\alpha\alpha}=0$,
we have omitted some irrelevant terms of order $q^4$, and we have set $\tau = 1
- T/T_c$. The fourth-order term included is the one responsible for
replica symmetry breaking in the SK model. The coefficients $w$ and $y$ are arbitrary
positive parameters. For the short-range KAS model, the bare propagator
is $g=1/(k^2-\tau)$.

To describe the long-range KAS model we replace the gradient terms in 
Eq.~(\ref{Hrep}) by
\begin{equation}
-\frac{1}{4} \sum_{\alpha, \beta} \int_{-L/2}^{L/2} dz\, \int_{-L/2}^{L/2} dz^{\prime}\,\frac{[q_{\alpha,\beta}(z)-q_{\alpha,\beta}(z^{\prime})]^2}{[(L/\pi) \sin (\pi (z-z^{\prime})/L)]^{2 \sigma}},
\label{eqn:longrange}
\end{equation} 
which on Fourier transforming can be seen to lead to a bare propagator
of the form $g=1/(k^{2 \sigma-1}- \tau)$ \cite{katzgraber:05c} as $k \to
0$. [Actually  Eq.~(\ref{eqn:longrange}) as it stands generates a
numerical factor of $c_g(\sigma)=-\Gamma(1-2 \sigma)\sin (\pi \sigma)$
in front of the $k^{2 \sigma-1}$ in the propagator, which can be removed
if desired by dividing Eq.~(\ref{eqn:longrange}) by $c_g(\sigma)$.]  In
terms of the original spins Eq.~(\ref{eqn:longrange}) is just
\begin{equation}
-\frac{1}{4} \sum_{\alpha, \beta} \sum_{i,j} {[J_{ij}^2]_{\rm av} \over (T_c^{MF})^2} 
(S_i^{(\alpha)} S_i ^{(\beta)}-S_j^{(\alpha)} S_j^{(\beta)})^2 \; .
\label{eqn:longrangespin}
\end{equation} 
The replica indices go $\alpha,\beta,\gamma=1, 2, \cdots, n, n+1,\cdots,
n+m$. The order parameter $q$ divides naturally into blocks of size $n$
and $m$. From now on, greek indices label the first block and roman ones
the second block, so, for example, $q_{\alpha a}$ means
$\alpha\in[1,n]$ and $a\in[n+1,n+m]$ and refers to the respective entry
in the off-diagonal or mixed sector.

Along the $z$-direction, which we take to be a distance along the
circumference of the ring of length $L$, we impose the boundary
condition that the solution is periodic in the Greek and Roman sectors,
and is antiperiodic in the mixed sectors reflecting the sign reversal of
the  bonds across the chosen diameter of the ring in the one sector with
respect to the other:
\begin{equation}
\begin{split}
q_{\alpha\beta}(z) &= 
  q_{\alpha\beta}(z+L)  \\
q_{ab}(z) &=
  q_{ab}(z+L)  \\
q_{\alpha a}(z) &=
  -q_{\alpha a}(z+L).
\end{split}
\label{bcAP}
\end{equation}
At mean-field level, there is the following \textit{stable} solution for $\ln
\overline{Z_{\rm P}^n Z_{\rm AP}^m}$:
\begin{align}
-\ln \overline{Z_{\rm P}^n Z_{\rm AP}^m} &= \beta\Hrep\{q^{\text{SP}}\},
\end{align}
where 
\begin{align}
q^{\text{SP}} &=
\left(\begin{array}{c|c}
Q^{(n)}\ab & 0 \\ \hline
0  & Q^{(m)}_{ij} 
\end{array}\right)
\label{eqn:SP}
\end{align} 
is independent  of the  spatial coordinates.  It  is natural  that the
diagonal  blocks are  the same  as the  regular Parisi  ansatz because
ordering in the system  with periodic boundary conditions, say, should
not be affected by there being another \textit{completely independent}
copy   with  different  boundary   conditions.   Choosing   the  mixed
greek-roman sector to vanish seems  to be consistent with the standard
interpretation \cite{marinari:00}  of   RSB  in  short-range  systems,
namely, that  changing  the  boundary conditions  changes  the  system
\textit{everywhere}.  More precisely  the surface  of the  domain wall
separating the regions that flip from the regions that do not flip is
space  filling. In  this  situation, one  can  reasonably expect  zero
overlap    between     configurations    with    different    boundary
conditions. However,  in the  droplet regime, where  there is  but one
state  and its  time  reversed,  we still expect that the thermal
average of the off-diagonal term remains zero. Our numerical work is
consistent with this assumption.

At  mean-field  level  the  solution  is  \textit{identical}  to  the
customary mean-field solution but for an $(n+m)$-times replicated system
($n+m$ being finite) \textit{without} boundary condition changes. We can
therefore immediately use  the result  from Ref.~\cite{aspelmeier:03}
that on the mean-field level, there is  no term of  order $(n+m)^2$, let
alone of order  $nm$, and thus the interface  energy vanishes to this
order.

We now turn to the loop expansion about the saddle point, which we  expect to
be valid for $\sigma <2/3$. The first correction is due to
Gaussian fluctuations around the saddle-point solution. They are given by
\begin{align}
-\ln \overline{Z_{\rm P}^n Z_{\rm AP}^m} &= \beta\Hrep\{q^{\text{SP}}\} +
  \frac 12 \sum_k I(k^{2\sigma-1}),
\end{align}
where
\begin{align}
    \label{Idef}
    I(k^{2\sigma-1}) &= \sum_{\mu}d_{\mu}\ln(k^{2\sigma-1}+\lambda_{\mu}).
\end{align}
Here $\lambda_{\mu}$ are the eigenvalues of the Hessian, evaluated at the
saddle-point solution and $d_{\mu}$ are their respective degeneracies.
These are the same as for a system of size $n+m$ without boundary
condition changes because the saddle-point solution is the same. Only
the nature of the $k$ vectors changes for the terms involving
eigenvalues whose corresponding eigenvectors $f$ are nonzero exclusively
in the mixed sector (i.e., $f\ab=f_{ij}=0$): The wave vectors have to
respect the imposed boundary conditions, which implies $k=(2n_d+1)\pi/L$
(with $n_d\in\mathbb{Z}$) in the mixed sector as opposed to
$k=2n_d\pi/L$ in the greek or roman sectors.

Following Refs.~\cite{aspelmeier:02} and \cite{aspelmeier:03}, it is
convenient to introduce the function
\begin{align}
\begin{split}
   & J(k^{2\sigma-1}):=  
     \ln(k^{2\sigma-1}+\frac{x_{1}^2w^2}{2y}) \\
&\quad - \frac{4w(4yk^{2\sigma-1} + wx_{1})}{4yk^{2\sigma-1}\sqrt{4yk^{2\sigma-1}+w^2x_{1}^2}}
     \tan^{-1}\frac{wx_{1}}{\sqrt{4yk^{2\sigma-1}+w^2x_{1}^2}},
\end{split}
\nonumber
\end{align}
where $x_1$ is the break-point of the Parisi $q$ function. This is
because the quadratic terms in $n$ and $m$ in $I$ are of the form
\begin{equation}
 \frac{(n+m)^2}{2}J_{\rm P}(k^{2\sigma-1}) +
nm[J_{\rm AP}(k^{2\sigma-1})-J_{\rm P}(k^{2\sigma-1})].
\nonumber
\end{equation}
The subscripts P and AP on $J$
mean that $J$ must be taken as $0$ when the argument is not of the required
type, i.e., periodic or antiperiodic.

We can now identify the term that gives rise to the interface free 
energy. Comparison with Eq.~\eqref{lnZnZm} shows
\begin{multline}
\label{ienergy}
\beta^2\overline{\DfPAP^2} = 
  \left({\sum}_{\rm AP}-{\sum}_{\rm P}\right) J(k^{2\sigma-1}) = \\
2 \sum_{r=1}^{\infty} 
\left[J\left(\left(\frac{(2r+1)\pi}{L}\right)^{2\sigma-1}\right) - J\left(\left(\frac{(2r)\pi}{L}\right)^{2 \sigma-1}\right)\right]+\\
\Delta f_{\mathrm{SK}}^2L^{2 \mu},
\end{multline}
where the subscripts on the sums indicate the nature of the allowed
$k$ vectors, as made explicit in the second part of Eq.~\eqref{ienergy}.
The sum over $k$ has been changed from $\pm\infty$ to $1$ to $\infty$
with the sum multiplied by a factor of $2$. The term $\Delta
f_{\text{SK}}^2L^{2 \mu}$ in Eq.~(\ref{ienergy}) comes from the $k = 0$
term in ${\sum}_{\rm P}$, which is nominally divergent as $k \rightarrow
0$.

In Ref.~\cite{aspelmeier:02} we  made an attempt at using finite-size
ideas to  regularize this divergence, but did  it incorrectly. It was
pointed   out,   correctly however,   that   the   diverging   term   is
identical to  the   variance  of   the  sample-to-sample
fluctuations  of   the  free  energy   of  the SK  model   containing
$L$ spins, $\Delta    f_{\mathrm{SK}}^2L^{2\mu}$, with   $\Delta
F_{\mathrm{SK}}$ an $L$ independent  term. Since that paper was written,
this variance  has    become   better   understood.    Parisi and
Rizzo \cite{parisi:10}     argued    that     $\mu=1/6$.  Aspelmeier
\cite{aspelmeier:08a,aspelmeier:08b} has shown   that
at least $\mu \le 1/4$. In Appendix \ref{sec:proof} the bound is
strengthened; $\mu \le 1/5$. We will take it that $\mu=1/6$.

Because $J(k^{2 \sigma-1})\approx-\pi w/4yk^{2 \sigma-1}$ for small
$k$, the term in the sum in Eq.~(\ref{ienergy}) is well approximated by
\begin{equation}
\frac{-\pi w}{4y}
2 \sum_{r=1}^{\infty}
\left[\frac{1}{\left[\frac{(2r+1)\pi}{L}\right]^{2 \sigma-1}} - 
  \frac{1}{\left[\frac{(2r)\pi}{L}] \right]^{2 \sigma-1}}\right]
= C L^{2 \sigma-1} ,
\nonumber
\end{equation}
where $C = [1-4^{-\sigma}(-4+4 ^{\sigma})\zeta(2 \sigma-1)]  \pi w/(2 \pi^{2 \sigma-1} y)$. This gives
\begin{equation}
\label{theta0}
\beta^2\overline{\DfPAP^2} = \Delta f_{\mathrm{SK}}^2L^{2 \mu} +C L^{2 \sigma-1}.
\end{equation}
Provided that $\mu=1/6$,  the right-hand  side  of  Eq.~(\ref{theta0})
is dominated by the first term. It is overtaken by  the second term only
when $\sigma >2/3$, but  when $\sigma >2/3$ one  is in  the droplet
region and   the  calculation   of   the   interface  free   energy
$\beta^2\overline{\DfPAP^2}$ takes a quite different form, as we will 
discuss below.

In the EA $d$-dimensional version of the calculation,  which was
summarized in Ref.~\cite{aspelmeier:08},  there was a similar change at
$d = 6$ dimensions. For the EA model the system is of length $L$ in the
$z$ direction, the direction in which the change is made from periodic
to antiperiodic boundary conditions, and it is periodic and of length
$M$ in the transverse $d-1$ dimensions, so $N=L M^{d-1}$. Then, for
$d>6$,
\begin{equation}
\beta^2\overline{\DfPAP^2}=  \Delta f_{\mathrm{SK}}^2N^{2 \mu}+L^2f(L/M).
\label{EAtheta}
\end{equation}
The term $\sim L^2 f(L/M)$ is the analog of the term $L^{2 \sigma-1}$
for the KAS model and is subdominant to the term of order $N^{1/3}$
(if $\mu=1/6$) until the dimensionality $d$ is lowered to $6$.  This
term depends on the shape of the system $L/M$ and has the aspect-ratio
scaling form expected for the interface free energy in dimensions $d \le
6$. The leading term in $N^{1/3}$ depends only on the total number of
spins $N$ and arises because the domain walls are space filling for $d
> 6$, with $d_s=d$. The interchange between the term in $N^{1/3}$ and
its leading correction is one of the reasons that we suspect that $6$ is
the dimension below which RSB behavior changes to droplet behavior. For
the KAS model, it is one of the reasons  why we believe that RSB
behavior does not occur in  the spin-glass phase for $\sigma \ge 2/3$.

\begin{figure}[htb]
\begin{center}
\includegraphics[width=\columnwidth]{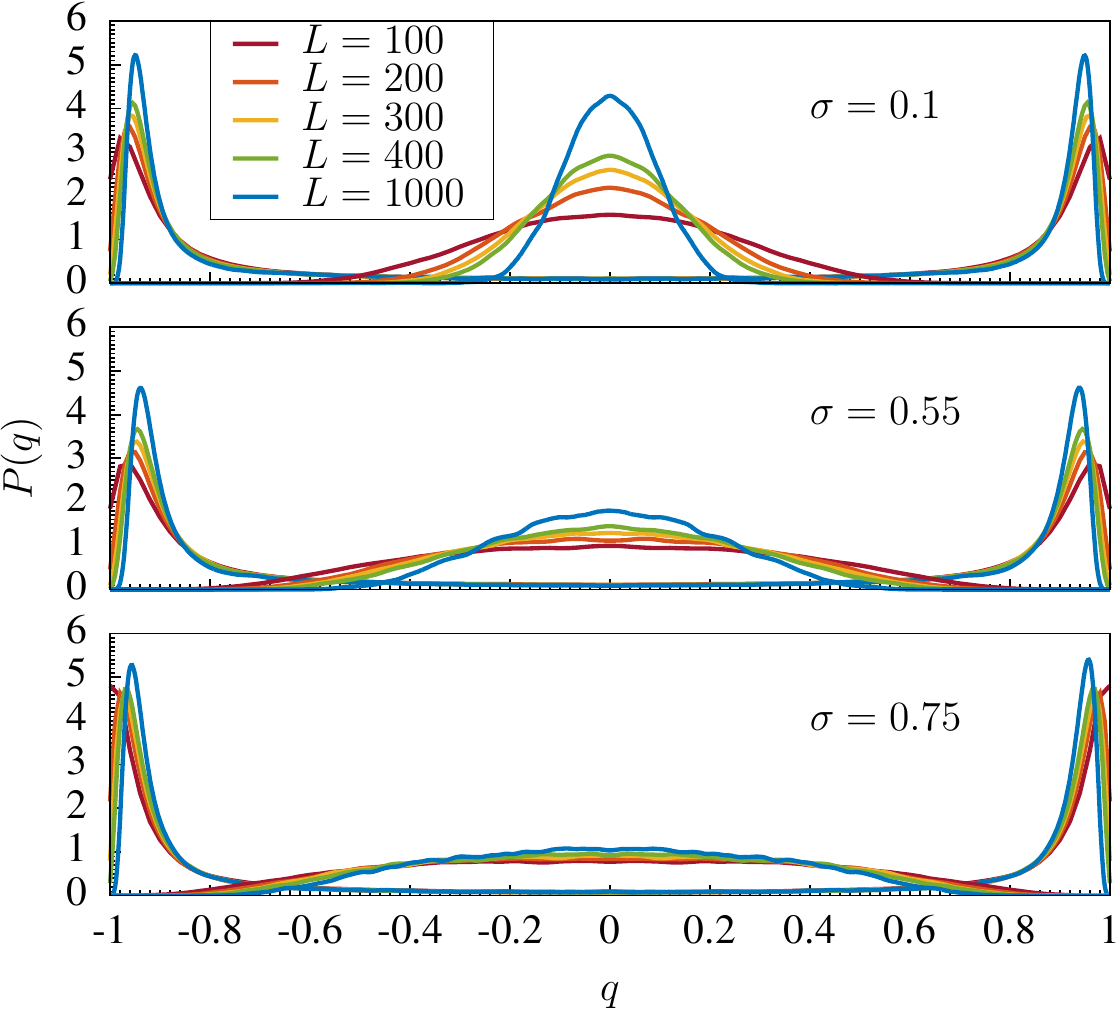}
\caption{(Color online)
Spin overlap distributions for three values of $\sigma$ as a function of
$L$ at $T=0.2T_c$. The distributions include $P^{\pbc,\pbc}(q)$,
$P^{\abc, \abc}(q)$, and $P^{\abc,\abc}(q)$. The diagonal
distributions have substantial peaks close to $\pm 1$, with decreasing
$q_{\rm{EA}}$ as $L$ increases, while the off-diagonal
distributions $P^{\pbc,\abc}(q)$ peak only at $q=0$, becoming increasingly
localized towards the center as $L$ increases for the system sizes
studied. Note that in the third panel, $P^{\pbc,\abc}(q)$ appears to
saturate to a non-$\delta$ function.  In all panels the systems sizes
increase from bottom to top as seen from the center of the distribution
for the cases where there is a central peak; otherwise, it is seen from the
peaks at large values of $|q|$.}
\label{Pq}
\end{center}
\end{figure}

\begin{figure}[htb]
\begin{center}
\includegraphics[width=\columnwidth]{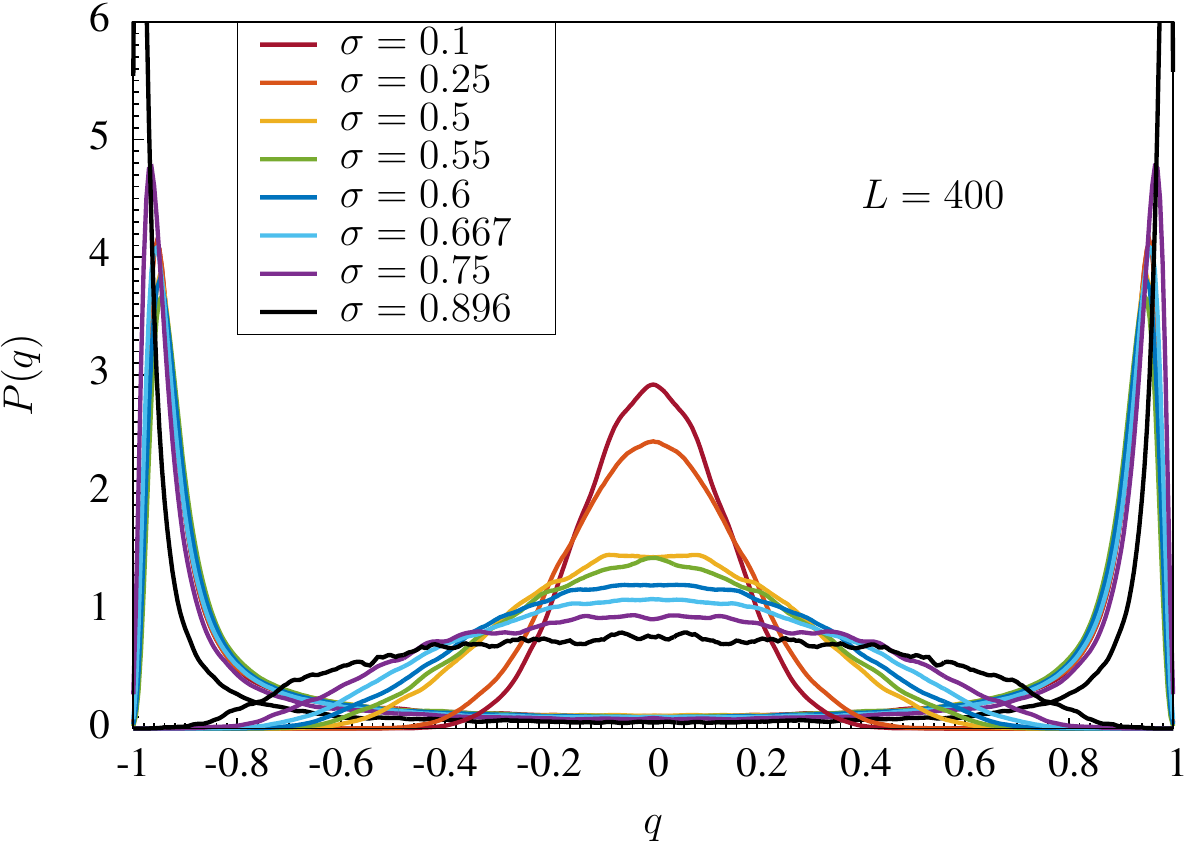}
\caption{(Color online)
Spin overlap distributions for $L=400$ for various values of $\sigma$ at
$T=0.2T_c$. The diagonal distributions are those with peaks close to
$\pm 1$, while  the off-diagonal distributions $P^{\pbc,\abc}(q)$
have peaks only at $q=0$ and become increasingly localized towards the
center as $\sigma$ decreases.  For the distributions with a peak at the
center, the values of $\sigma$ increase with decreasing peak height. For
the distributions with large support for $|q|$ large the values of
$\sigma$ increase for increasing peak height.}
\label{Pq400}
\end{center}
\end{figure}

The key assumption used in our calculation is that in the greek-roman
sector $Q_{\alpha a}= \langle q_{\alpha a}(z) \rangle = 0$. This
assumption allowed us to expand about a spatially uniform solution. In
Appendix \ref{sec:numerical} we give the numerical details of the
simulations that were done to directly test this assumption. We study
the three overlap functions $P^{\pbc,\pbc}(q)$, $P^{\pbc, \abc}(q)$, and
$P^{\abc,\abc}(q)$. Thus, the overlap $q$ between the spin
$S_i^{(\pbc)}$ in the system with periodic boundary conditions and the
spin $S_i^{(\abc)}$ at the same site $i$ in the system with
antiperiodic boundary conditions is defined as
\begin{equation}
q =\frac{1}{L} \sum_i^{L} S_i^{(\pbc)} S_i^{(\abc)}.
\label{eqn:pqdef}
\end{equation}
The distribution of this overlap is $P^{\pbc, \abc}(q)$, and together with 
 the similarly-defined overlap distributions $P^{\pbc, \pbc}(q)$ and $P^{\abc,
\abc}(q)$ is shown in Figs.~\ref{Pq} and \ref{Pq400} for a variety of
system sizes $L$ and $\sigma$ values. We refer to the last two
distributions as the diagonal contributions [after bond averaging
$P^{\pbc, \pbc}(q)=P^{\abc, \abc}(q)$] and $P^{\pbc, \abc}(q)$ as the
off-diagonal contribution. In replica language, the overlap defined
in Eq.~(\ref{eqn:pqdef}) relates to that in the mixed greek-roman
sector $q_{\alpha a}$. Our crucial assumption was that $Q_{\alpha
a}(z)= \langle q_{\alpha a}(z) \rangle =0$. One might have expected that
in the mixed sector $Q_{\alpha a}(z)$ is an odd function interpolating
at one end of the system from $+q_{EA}$ to $-q_{EA}$ at the other end in
order to satisfy the boundary conditions. However, if that were the
situation, the off-diagonal distribution $P^{\pbc, \abc}(q)$ would have
peaks near $\pm q_{EA}$, just like the peaks of the diagonal
distributions. However, the only peak in the off-diagonal distribution
occurs at $q=0$ and for all values of $\sigma$ there are no signs of
peaks at $\pm q_{EA}$. We believe that this confirms our fundamental
assumption.

We find it useful to examine the second moment of $P^{\pbc,
\abc}(q)$, which equals $\overline{\langle q^2 \rangle}$, where
\begin{equation}
 q^2=\frac{1}{L^2} \sum_{j,i}  S_j^{({\pbc})} S_j^{({\abc})} S_i^{({\pbc})} S_i^{({\abc})}.
\end{equation}
Let us examine the situation at zero temperature.  Let
$\tau_i=S_i^{({\pbc})} S_i^{({\abc})} =\pm 1$. Then $\tau_i=+1$ if at site
$i$ the spins associated with periodic and antiperiodic boundary
conditions are parallel; $\tau_i=-1$, if these spins are antiparallel.
A sequence in which the $\tau_i$ are of the same sign will be called an
island. Then
\begin{equation}
q^2=\left(\frac{1}{L} \sum_i \tau_i\right)^2 .
\label{eqn:q2def}
\end{equation}
For  the one-dimensional  KAS  model with  long-range interactions,  a
droplet may consist of disconnected pieces, i.e., islands, so a fractal
dimension $d_s$  could be defined if  the number of  islands scales as
$L^{d_s}$. In the  RSB region  we expect  that $d_s=d  =1$.   If one
changes  the boundary  conditions from  periodic to  antiperiodic, one
does not  generate a  single reversed domain  but instead a  number of
order $L^{d_s}$ islands. The islands have  a  distribution of sizes.  In
the  RSB region ($\sigma < 2/3$) we expect that the number of these
islands varies  as $L/L_0$, where $L_0(\sigma)$ is the root-mean-square
size  of the islands,  which seems to  increase with $\sigma$. This
break-up into islands arises to  reduce the energy by taking advantage
of particular features  of the bonds $J_{ij}$ and the existence of many
states in  the RSB region.  Because islands are only a feature  of
long-range one-dimensional  systems, they have  not been studied  in
the  literature. In  the  EA  model  with  short-range interactions, the
droplets  are simply connected.

The first moment of $P(q)$ equals $\overline{\langle  q \rangle}$ and is
zero [the functions are symmetric, so $P(q)=P(-q)$].  Thus, the average
value of $\tau_i$ is zero and there are as many positive $\tau_i$ values
as negative $\tau_i$  values.  For any  given ground state, the  average
of $\tau_i$ might not be zero. However, if one  averages over the ground
state and the states obtained by flipping all the spins in (say) the
system  with  periodic  boundary  conditions,  the  average  value  of
$\tau_i$ will be zero.

\begin{figure}[htb]
\begin{center}
\includegraphics[width=\columnwidth]{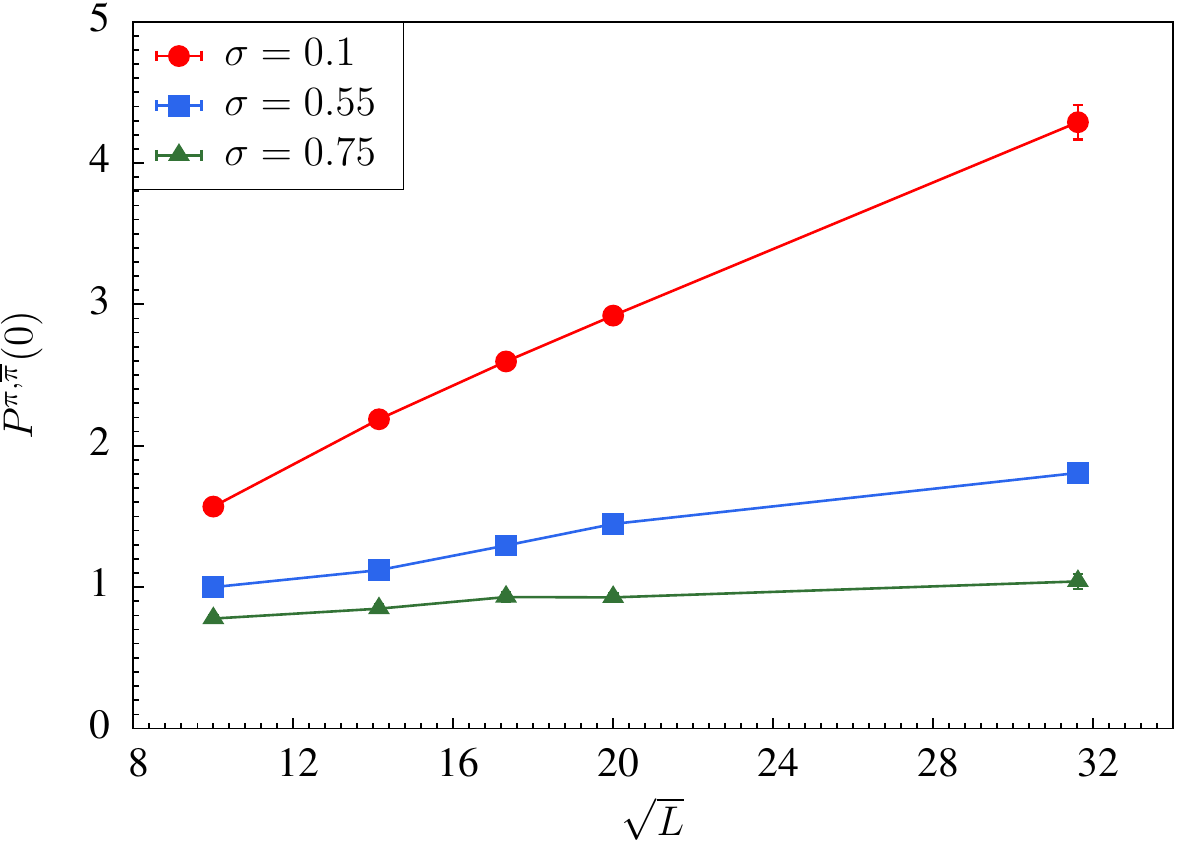}
\caption{(Color online)
Parisi overlap $P^{\pbc,\abc}(0)$ as a function of $\sqrt{L}$, for three
representative values of $\sigma$ at $T=0.2T_c$. Note that
$P^{\pbc,\abc}(0)$ grows approximately linearly in $\sqrt{L}$ in the RSB
regime, but seems to level off in the droplet or scaling regime ($\sigma
>2/3$).
}
\label{Pq3}
\end{center}
\end{figure}

The second moment can be estimated by noting that the sum in
Eq.~(\ref{eqn:q2def}) is a sum of $L/L_0$ terms random in sign and of
magnitude $L_0$, so the sum is of order $\sqrt{L/L_0} \,L_0$. Hence,
$\overline{q^2} =L_0/L$. Assuming that the distribution of $q$ is
Gaussian,
\begin{equation}
P^{\pbc, \abc}(q)= \sqrt{\frac{L}{2 \pi L_0}}\exp\left[-\frac{L q^2}{2 L_0} \right].
\label{eqn:SKPq}
\end{equation}
Thus, in the limit of $L \to \infty$, $P^{\pbc, \abc}(q) =\delta(q)$.
The peak $P^{\pbc, \abc}(0)$ is expected to vary as $\sim \sqrt{L/L_0}$.
It is shown in Fig.~\ref{Pq3} and  seems to be consistent with these
arguments at least for the data for $\sigma =0.1$ and $0.55$, which lie
in the RSB region.

In  the droplet  region the  data  in Figs.~\ref{Pq} and  \ref{Pq400}
imply that $\overline{\langle q^2 \rangle}$  is nonzero as $L\to
\infty$. Again, $P^{\pbc,\abc}(q)$ is a  function of $q$ centered at the
origin and of nonzero width, so  the peak $P^{\pbc,\abc}(0)$ remains
finite in the droplet region. Therefore, there seems to be a simple test
for determining whether the system has RSB behavior  or not.  If there
is RSB behavior,  $P^{\pbc, \abc}(0)$ diverges with the  system size,
whereas in the  droplet region  it stays finite.  Simulations of the
three-dimensional EA  model suggest that it stays  finite \cite{wang:16x}.
Our  numerical work shows that in the KAS  model the change from RSB  to
droplet behavior might  occur  somewhere between $\sigma =0.55$ and
$\sigma  =0.75$, but finite-size effects  make it hard to pin down the
change more precisely and we have failed to find any method of analysis
that even hints at a sharp feature at $\sigma =2/3$. It might  be that
$L_0$ diverges as $\sigma \to 2/3$, so that $\overline{\langle q^2
\rangle}$  joins smoothly  to  its expected finite form for $\sigma \ge
2/3$.  We tried  to determine whether $L_0$ has this feature, but failed
to see it clearly, probably because of finite-size issues.  We do
emphasize, however, that the window $0.55 \le \sigma \le 0.75$
corresponds for a hypercubic system to space dimensions between
approximately $4$ and $10$.

In the  RSB region the loop expansion,  i.e., the expansion  about the
mean-field  solution, is  well controlled (but technically challenging).
Unfortunately, such a perturbative  approach completely fails  in the
droplet  region as  the  terms in  the  expansion about  the state  of
assumed  replica  symmetry  appear  to break  replica  symmetry.  This
problem  might be overcome  by going  to all  orders in  the expansion
\cite{moore:05}. However, we can get the exponent $\theta$ within our
formalism by using Eq.~(\ref{eqn:longrangespin}) for the bending energy
and using the arguments in Refs.~\cite{fisher:88,bray:86,moore:10}. It is
useful to set $\tau_i^{\alpha a}=S_i^{(\alpha)} S_i^{(a)}=\pm 1$, so
that $\tau_i^{\alpha a}=+1$ if the spins $S_i^{(\alpha})$ and
$S_i^{(a)}$ are parallel and $-1$ otherwise. Then, by flipping (say)
half the $\tau_i^{\alpha a}$ spins, one can see that the variance of the
replicated bending energy scales as $mn L^{2-2 \sigma}$, just as already
argued in Refs.~\cite{fisher:88,bray:86,moore:10}. In that case
\begin{equation} 
\theta =1-\sigma.
\label{eqn:thetadrop}
\end{equation}
We believe that Eq.~(\ref{eqn:thetadrop})  applies only  in the  droplet 
region, i.e., $ \sigma \ge 2/3$.  However, in  the  region $\sigma  <2/3$ where  we
expect $\overline{\langle q^2 \rangle}$ to be of order $L_0/L$,  the  presence of
so many islands (of order $L/L_0$) of finite size $L_0$ and the correlations between them must allow the  system to reduce  the free-energy variance
associated with the  transition   from  periodic   to antiperiodic
boundary conditions from this estimate of $L^{2(1-\sigma)}$ to the smaller value of $L^{1/3}$.

For $\sigma \ge 1$, the exponent $\theta$ is no longer positive and
there   will    be   no   finite-temperature    spin-glass   phase
\cite{moore:10}. However, the short-range EA model value for $\theta$ is
$-1$ \cite{bray:86} and  so  the crossover  to  the  short-range
behavior occurs   above  $\sigma =2$ when   the  long-range interactions
become irrelevant at the  zero-temperature  fixed point
\cite{bray:86b,fisher:88}.

\begin{figure}[htb]
\begin{center}
\includegraphics[width=\columnwidth]{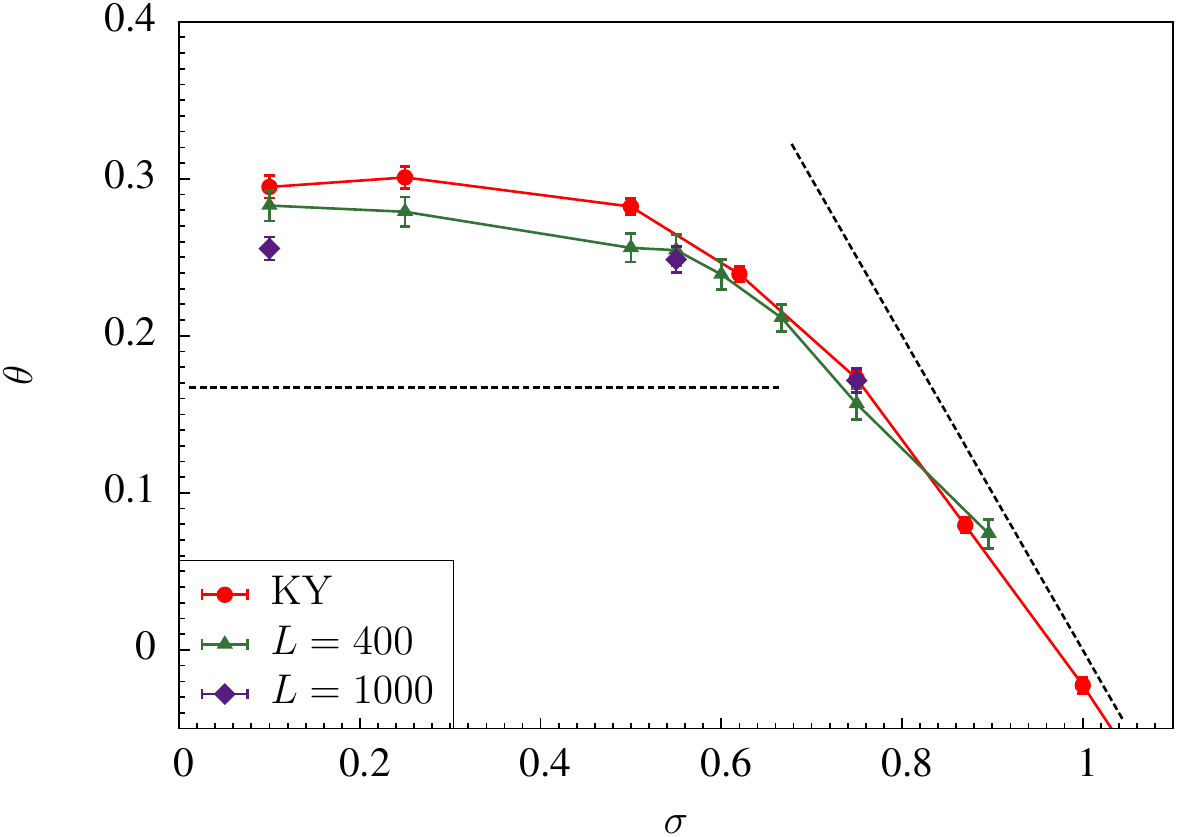}
\caption{(Color online) 
Estimates of the exponent $\theta$ as a function of $\sigma$. Here KY denotes
results obtained on samples up to $L =256$ in Ref.~\cite{katzgraber:03}
by Katzgraber and Young. The dashed line denotes the droplet regime
prediction for $\theta=1-\sigma$. We expect this to apply for
 $2> \sigma \ge 2/3$. When $\sigma <2/3$ we predict that $\theta=1/6$
and the horizontal dashed line shows this prediction. Notice that the
result for $\theta$ in the SK model region ($\sigma =0.1$) is moving
closer to the theoretical prediction of $1/6$ as $L$ increases, albeit
very slowly.
}
\label{theta}
\end{center}
\end{figure}

\section{Conclusions}
\label{sec:conclusions}

We have predicted for the one-dimensional KAS  model that in the RSB region
($\sigma < 2/3$) $\theta  =1/6$, while in the region $2/3 \le \sigma
<2$, $\theta =(1-\sigma)$. Notice that at the borderline of  the RSB
region and droplet region at $\sigma =2/3$, $\theta$ is predicted to be
discontinuous, as shown in Fig.~\ref{theta}.

This discontinuity seems to be a feature of the KAS model only. For the
$d$-dimensional EA model where six is the borderline dimension, there is
evidence that $\theta$ is continuous at six dimensions as it approaches
unity in six dimensions (see Refs.~\cite{boettcher:05d,boettcher:04b}
for numerical evidence  on this question). If it tends to unity
approaching six dimensions from below, it merges with the value of
$\theta$ expected from RSB as the dimension $d$ approaches six from
above, as given in Eq.~(\ref{EAtheta}). In addition, $\theta$ and $\mu$
have been studied as a function of $\sigma$ via numerical simulations.
This was first done by Katzgraber and Young (KY)
\cite{katzgraber:03,katzgraber:03f}, with results  that are not very
close to the predictions made here. No  discontinuity in $\theta$  was
reported  at $\sigma =2/3$.   We  believe  that  the discrepancies are
due to  finite-size effects \cite{aspelmeier:16}, which are surprisingly
large in the KAS model. The  data produced in the  present study  allows
us  to reach larger  sizes  than  those previously studied by KY, who
studied  $L \le 256$. The larger sizes that we studied, $L=400$ and
$L=1000$,  do give results somewhat  closer to  our theoretical
expectations,  but the  movement towards them is slow. In the  droplet
region the  finite-size effects are probably of the same origin as those
that make the Parisi overlap $P^{\pbc,\pbc}(q=0)$ nonzero, contrary to
the arguments  of droplet theory, i.e., the system sizes studied  are
just not  large enough  to  make it vanish.  Smaller systems  appear to
have RSB  features such  as a nonzero value of $P^{\pbc,\pbc}(0)$.

In  the RSB  region where $\sigma  <2/3$, we  predict that $\theta
=1/6$. The  value of  $1/6$ is  the SK value for $\mu$.  However, the
values for $\mu$ mostly reported in the numerical literature
\cite{boettcher:05,katzgraber:05d,boettcher:10} for the SK model seem
closer to a value around $0.25$, which, while very different from $1/6$
of the theoretical work of Parisi and Rizzo \cite{parisi:10}, is
consistent with the numerical value for $\theta$ reported in
\cite{katzgraber:03}.  However, once again, we  suspect that
finite-size effects  in  the RSB  region might cause  the discrepancy.
In Appendix \ref{sec:proof} we give what we believe is a cogent argument
that at least $\mu \le 1/5$.

Our  work suggests that a convenient numerical test for RSB  or droplet
behavior is via the size dependence of $P^{\pbc,\abc}(0)$. If this
quantity does not grow with system size, the ordered state is droplet
like. If it grows with system size, the system has RSB behavior.
However, this test is affected by finite-size effects, yet perhaps not
as badly as other commonly used tests based on the existence or not of
the AT line. Simulations using special-purpose machines
\cite{belletti:08} that allow for considerably larger system sizes might
allow for the detection of the true nature of the spin-glass state using
the metric introduced herein.

\begin{acknowledgments}

We would like to thank Stefan Boettcher for bringing us up to date on
his latest numerical work and Jon Machta for a discussion regarding
$P^{\pbc,\abc}(0)$ as a function of the system size $L$.  T.A.~would
like to thank Christoph Norrenbrock for useful discussions.  W.W.~and
H.G.K.~acknowledge support from the National Science Foundation (Grant
No.~DMR-1151387). H.G.K.~thanks P.~Hobbs for providing multiple sources
of inspiration. The work of H.G.K.~and W.W~was supported in part by the
Office of the Director of National Intelligence (ODNI), Intelligence
Advanced Research Projects Activity (IARPA), via MIT Lincoln Laboratory
Air Force Contract No.~FA8721-05-C-0002.  The views and conclusions
contained herein are those of the authors and should not be interpreted
as necessarily representing the official policies or endorsements,
either expressed or implied, of ODNI, IARPA, or the U.S.~Government. The
U.S.~Government is authorized to reproduce and distribute reprints for
Governmental purpose notwithstanding any copyright annotation thereon.
We thank Texas A\&M University for access to their Ada and Curie
clusters.

\end{acknowledgments}

\appendix
\section{``Proof'' that $\mu\le\frac 15$ for the SK model}
\label{sec:proof}

For $\sigma < 2/3$, our calculation of the exponent $\theta$ related it
to the exponent $\mu$ of the sample-to-sample variation of the energy in
the SK limit. This exponent is believed to be $1/6$ \cite{parisi:10},
but numerical studies of it give larger values
\cite{katzgraber:03,boettcher:05,katzgraber:05d,boettcher:10}. In this
appendix we derive an upper bound on its value, namely, $\mu \le 1/5$.
We believe that with the methods used here it might be possible
eventually to actually prove that $\mu = 1/6$.  We also point out that
the numerical work is done for the ground state, i.e., the the free
energy at $T = 0$, and the argument in this appendix is for the free
energy at a finite temperature $T< T_c$.  However, we do not think this
difference affects the value of $\mu$. The difference between the
numerical value and our theoretical expectations, is, we believe, just
another problem caused by finite-size effects.

In Refs.~\cite{aspelmeier:08a,aspelmeier:08b} it was shown that the
free-energy fluctuations $\Delta F$ in the SK model are given by the
exact formula
\begin{align}
\beta^2 \Delta F^2 &= \frac{N^2\beta^4}{16}\int_0^\infty f_2(\epsilon)\mathbb E \langle (q_{13}^2-q_{14}^2)(q_{13}^2-q_{23}^2) \rangle \, d\epsilon \nonumber \\
&+ \frac{N\beta^2}{4} \int_0^\infty g_2(\epsilon) \left(\mathbb E\langle q_{13}^2\rangle -\frac 1N\right)\, d\epsilon,
\label{fluct}
\end{align}
where $N$ is the system size, $\beta$ is the inverse temperature, $f_2$ 
and $g_2$ are two functions defined by
\begin{align*}
f_2(\epsilon) &= \frac{2\epsilon \ln(1+\epsilon^2)}{(1+\epsilon^2)^2} &
g_2(\epsilon) &=\frac{\epsilon \ln(1+\epsilon^2)}{(1+\epsilon^2)^{3/2}},
\end{align*}
and $q_{ij}$, with $i=1$, $2$ and $j=3$, $4$, are the overlaps
between spin-glass systems $1,\dots,4$ of which systems $1$ and $2$ have
identical Gaussian bonds $J^{(i)}_{kl}$ with unit variance, and likewise
for systems $3$ and $4$ with bonds  $J^{(j)}_{mn}$, and the correlation
between the two sets of bonds is given for $k>l$ and $m>n$ by
\begin{align*}
\mathbb E J^{(i)}_{kl}J^{(j)}_{mn} &= \delta_{km}\delta_{ln} \frac{1}{\sqrt{1+\epsilon^2}}.
\end{align*}
The symbol $\mathbb E$ here stands for the expectation value with
respect to all bonds and the angular brackets denote a thermal average.
The free-energy fluctuations are thus directly linked to bond chaos via
integrals over a function ($f_2$ or $g_2$) times momenta of overlaps
between spin-glass replicas with different but correlated bonds.

For the calculation of Eq.~\eqref{fluct} it is, in principle, necessary
to calculate 3- and 4-replica overlaps of the form $\mathbb E \langle
q_{13}^2q_{14}^2\rangle$, etc. This is, however, very difficult.
Instead, we note that trivially
\begin{align*}
0 &\le (q_{14}^2-q_{23}^2)^2 = q_{14}^4 + q_{23}^4 - 2q_{14}^2q_{23}^2,
\end{align*}
whence it follows that
\begin{align*}
\mathbb E\langle q_{14}^2q_{23}^2\rangle &\le \mathbb E\langle q_{13}^4\rangle,
\end{align*}
since replicas $1$ and $2$ are identical, as are replicas $3$ and $4$,
and so $\mathbb E\langle q_{14}^4\rangle = \mathbb E\langle
q_{23}^4\rangle = \mathbb E\langle q_{13}^4\rangle$. This implies that
\begin{multline}
 \mathbb  E\langle(q_{13}^2-q_{14}^2)(q_{13}^2-q_{23}^2) \rangle = 
 \\  \mathbb E \langle q_{13}^4 -
q_{13}^2q_{23}^2 - q_{14}^2q_{13}^2 + q_{14}^2q_{23}^2\rangle \le 2\mathbb E\langle q_{13}^4\rangle.
\label{bound}
\end{multline}
For an upper bound of the first integral term in Eq.~\eqref{fluct} it is
therefore only necessary to know $\mathbb E\langle q_{13}^4\rangle$ as a
function of $\epsilon$. Such moments have been calculated asymptotically
in various regimes in Ref.~\cite{aspelmeier:08a}. The results are summarized
in Table \ref{lala}.

\begin{table}
\caption{
Summary of moments $\mathbb E\langle q_{13}^k\rangle$ with $k = 2$, $4$
calculated in Ref.~\cite{aspelmeier:08a}.
\label{lala}
}
\begin{tabular*}{\columnwidth}{@{\extracolsep{\fill}} l c c c}
\hline
\hline
& Regime I & Regime II & Regime III \\
\hline
& $\epsilon\ll N^{-1/2}$ & $N^{-1/2}\ll\epsilon\ll N^{-1/5}$& $N^{-1/5}\ll\epsilon$ \\\hline
$\mathbb E\langle q_{13}^2\rangle$ & const & $\sim (N\epsilon^2)^{-2/3}$ & $\sim [Nh(\epsilon)]^{-1}$\\
$\mathbb E\langle q_{13}^4\rangle$ & const & $\sim (N\epsilon^2)^{-4/3}$ & $\sim [Nh(\epsilon)]^{-2}$\\
\hline
\hline
\end{tabular*}
\end{table}

The function $h$ is a nonnegative function with the
features that $h(\epsilon)= O(\epsilon^3)$ for $\epsilon\to 0$
and $h(\epsilon)\to\text{const}$ for $\epsilon\to\infty$. These results
allow for calculating the asymptotic behavior of the integrals in
Eq.~\eqref{fluct}. The first integral can, with the help of
Eq.~\eqref{bound}, be bounded by
\begin{align*}
\frac{N^2\beta^4}{16}\int_0^\infty f_2(\epsilon)\mathbb E \langle (q_{13}^2-q_{14}^2)(q_{13}^2-q_{23}^2) \rangle \, d\epsilon & \nonumber \\ \le \frac{N^2\beta^4}{8}\int_0^\infty f_2(\epsilon)\mathbb E \langle q_{13}^4\rangle\, d\epsilon \\
= \frac{N^2\beta^4}{8}\int_0^{N^{-1/5}} \epsilon^3 \mathcal F(N^{1/2}\epsilon)\,d\epsilon \\
+ \frac{N^2\beta^4}{8}\int_{N^{-1/5}}^{\epsilon_0} \epsilon^3 (N\epsilon^3)^{-2}\,d\epsilon
+ O(1),
\end{align*}
where $\mathcal F$ is a scaling function combining regimes I and II and
with the properties $\mathcal F(x)\to\text{const}$ as $x\to 0$ and
$\mathcal F(x) = O(x^{-8/3})$ as $x\to\infty$. The term
$\epsilon^3$ in the integrals comes from a Taylor expansion of $f_2$
for small $\epsilon$. The upper limit of the second part of the integral, which corresponds to regime III, is some fixed $\epsilon_0$ of order 1 but small enough to allow for a Taylor expansion of $f_2$ and $h$. Asymptotic evaluation of the integral is now
possible and the result is, for regimes I and II,
\begin{align*}
\frac{N^2\beta^4}{8}\int_0^{N^{-1/5}} \epsilon^3 \mathcal F(N^{1/2}\epsilon)\,d\epsilon & \nonumber \\=
\frac{\beta^4}{8} \int_0^{N^{3/10}} x^3\mathcal F(x)\,dx \sim  N^{2/5}.
\end{align*}
The dominant contribution to regime III of the integral comes from the lower bound and is also $\sim N^{2/5}$.
A similar calculation shows that the second integral term in
Eq.~\eqref{fluct} is subdominant to $N^{2/5}$, hence the fluctuations
are bounded by
\begin{align*}
\beta^2 \Delta F^2 \le \text{const}\times N^{2/5}
\end{align*}
and the fluctuation exponent $\mu$ in $\beta \Delta F \sim N^{\mu}$ is
bounded by $\mu \le 1/5$.

\section{Numerical simulation details}
\label{sec:numerical}

The main purpose of the numerical work is  to verify the main assumption
in our calculation in  Sec.~\ref{sec:theta}.  This is that in the mixed
sector $Q_{\alpha a}=\langle q_{\alpha  i}(z)\rangle=0$.  It  is this
assumption that allowed  us to  construct the first  term in the  loop
expansion about a spatially uniform solution. We also expect that
$\langle q_{\alpha  i}(z)\rangle=0$ in the droplet region.  Our studies
of the exponents $\theta$ and $\mu$ are in effect a by-product of these
investigations.

When doing numerical work on the one-dimensional long-range model, one
has  to  decide whether  to  stay with  the  KAS  model as  originally
outlined, in which  every spin is coupled to every  other spin, or the
diluted  model in  which only  a fixed  number $z$  (typically  $z$ is
chosen  to be  $6$) of  the  spins are  coupled
\cite{leuzzi:08,katzgraber:09b}.  The advantage  of the  diluted model
is that the  simulations are faster, because each spin update requires
only a constant number of updates from their neighbors.  On the other hand,
there is  a consequence in that it suffers from larger finite-size
effects. We therefore decided to study the fully connected model.
Despite smaller system sizes than in the diluted case, finite-size
corrections to scaling are smaller.

\begin{table}
\caption{
Parameters of the simulations for different values of $\sigma$ and
system size $L$ for periodic and antiperiodic
antiperiodic boundary conditions. Here $R_0$ is the population size, $T_0 =
1/\beta_0$ is the lowest temperature simulated, $N_T$ is the number of
temperatures used in the annealing schedule, and $M$ is the number of
disorder realizations.
\label{table}
}
\begin{tabular*}{\columnwidth}{@{\extracolsep{\fill}} c c c c c c}
\hline
\hline
$L$ &$\sigma$  & $R_0$ & $1/\beta_0$ & $N_T$  & $M$ \\
\hline
$100$ &\{0.1,0.25,0.5,0.55\}  & $ 10^4$ & $0.1000$  & $101$ &$6000$ \\
$100$ &\{0.6\}  & $ 10^4$ & $0.0934$     & $101$   & $6000$ \\
$100$ &\{0.667\}  & $ 10^4$ & $0.0833$     & $101$   & $6000$ \\
$100$ &\{0.75\}  & $ 10^4$ & $0.0690$     & $101$   & $6000$ \\
$100$ &\{0.896\}  & $2\times10^4$ & $0.0373$     & $101$   & $12000$ \\

$200$ &\{0.1,0.25,0.5,0.55\} & $2\times10^4$ & $0.1000$ & $101$ &$6000$ \\
$200$ &\{0.6\}  & $2\times10^4$ & $0.0934$     & $101$   & $6000$ \\
$200$ &\{0.667\}  & $2\times10^4$ & $0.0833$     & $101$   & $6000$ \\
$200$ &\{0.75\}  & $2\times10^4$ & $0.0690$     & $101$   & $6000$ \\
$200$ &\{0.896\}  & $2\times10^4$ & $0.0373$     & $101$   & $6000$ \\

$300$ &\{0.1,0.25,0.5,0.55\}  &$4\times10^4$ & $0.1000$ & $101$ &$6000$ \\
$300$ &\{0.6\}  & $4\times10^4$ & $0.0934$     & $101$   & $6000$ \\
$300$ &\{0.667\}  & $4\times10^4$ & $0.0833$     & $101$   & $6000$ \\
$300$ &\{0.75\}  & $4\times10^4$ & $0.0690$     & $101$   & $6000$ \\
$300$ &\{0.896\}  & $4\times10^4$ & $0.0373$     & $101$   & $6000$ \\

$400$ &\{0.1,0.25,0.5,0.55\}  &$5\times10^4$ & $0.1000$ & $201$ &$6000$ \\
$400$ &\{0.6\}  & $5\times10^4$ & $0.0934$     & $201$   & $6000$ \\
$400$ &\{0.667\}  & $5\times10^4$ & $0.0833$     & $201$   & $6000$ \\
$400$ &\{0.75\}  & $5\times10^4$ & $0.0690$     & $201$   & $6000$ \\
$400$ &\{0.896\}  & $5\times10^4$ & $0.0373$     & $201$   & $6000$ \\

$1000$ &\{0.1,0.55\}  &$2\times10^5$ & $0.1000$ & $201$ &$3000$ \\
$1000$ &\{0.75\}  &$2\times10^5$ & $0.0690$ & $201$ &$3000$ \\
\hline
\hline
\end{tabular*}
\label{tab:simdetails}
\end{table}

\begin{table}
\caption{
Dependence of $T_c(\sigma)$ on $\sigma$. The values of $T_c$ used in the
simulation and the error bars are estimated using
the data of Ref.~\cite{katzgraber:05c} via a cubic spline interpolation.
}
\begin{tabular*}{\columnwidth}{@{\extracolsep{\fill}} l r}
\hline
\hline
$\sigma$ & $T_c(\sigma)$ \\
\hline
$0.55  $  & $1.00(3)$ \\
$0.6   $  & $0.93(3)$ \\
$0.6667$  & $0.83(2)$ \\
$0.75  $  & $0.69(1)$ \\
$0.896 $  & $0.37(1)$ \\
\hline
\hline
\end{tabular*}
\label{table:Tc}
\end{table}

The model is simulated using the population annealing Monte
Carlo  method\cite{hukushima:03,zhou:10,machta:10,wang:15e}. Population
annealing works with a large population $R_0$ of replicas of the system,
each with the same disorder. The population transverses an annealing
schedule and maintains thermal equilibrium to a low target temperature
$T_0 = 1/\beta_0$. In this work we used a schedule that is linear in
$\beta$. When the temperature is lowered from $\beta$ to $\beta^\prime$
the population is resampled. The mean number of copies of replica $i$ is
proportional to the appropriate reweighting factor
$\exp[-(\beta^\prime-\beta) E_i]$. The constant of proportionality is
chosen such that  the population size remains close to $R_0$. This is
followed by $N_S=10$ sweeps of the Metropolis Monte Carlo algorithm of
each  replica.  We  simulate $M$ disorder realizations  and  measure
overlaps at $T=T_0=0.1T_c$ and $T=0.2T_c$.  The  simulation parameters
are summarized in Table~\ref{tab:simdetails}. Our estimates of
$T_c(\sigma)$ are given in Table~\ref{table:Tc}. Most of our studies of
the three overlap functions were done at $0.2 T_c(\sigma)$, in order to
more easily compare how varying $\sigma$ affects them. We find the
ground-state energy by finding the lowest energy in our population at
the lowest temperature and we ensure that the number of replicas having the
lowest energy is large, in order to  estimate the exponents
$\theta$ and $\mu$.

\bibliography{refs}

\begin{thebibliography}{52}
\expandafter\ifx\csname natexlab\endcsname\relax\def\natexlab#1{#1}\fi
\expandafter\ifx\csname bibnamefont\endcsname\relax
  \def\bibnamefont#1{#1}\fi
\expandafter\ifx\csname bibfnamefont\endcsname\relax
  \def\bibfnamefont#1{#1}\fi
\expandafter\ifx\csname citenamefont\endcsname\relax
  \def\citenamefont#1{#1}\fi
\expandafter\ifx\csname url\endcsname\relax
  \def\url#1{\texttt{#1}}\fi
\expandafter\ifx\csname urlprefix\endcsname\relax\def\urlprefix{URL }\fi
\providecommand{\bibinfo}[2]{#2}
\providecommand{\eprint}[2][]{\url{#2}}

\bibitem[{\citenamefont{Edwards and Anderson}(1975)}]{edwards:75}
\bibinfo{author}{\bibfnamefont{S.~F.} \bibnamefont{Edwards}} \bibnamefont{and}
  \bibinfo{author}{\bibfnamefont{P.~W.} \bibnamefont{Anderson}},
  \emph{\bibinfo{title}{Theory of spin glasses}}, \bibinfo{journal}{J. Phys. F:
  Met. Phys.} \textbf{\bibinfo{volume}{5}}, \bibinfo{pages}{965}
  (\bibinfo{year}{1975}).

\bibitem[{\citenamefont{Parisi}(1979)}]{parisi:79}
\bibinfo{author}{\bibfnamefont{G.}~\bibnamefont{Parisi}},
  \emph{\bibinfo{title}{Infinite number of order parameters for spin-glasses}},
  \bibinfo{journal}{Phys. Rev. Lett.} \textbf{\bibinfo{volume}{43}},
  \bibinfo{pages}{1754} (\bibinfo{year}{1979}).

\bibitem[{\citenamefont{Parisi}(1980)}]{parisi:80}
\bibinfo{author}{\bibfnamefont{G.}~\bibnamefont{Parisi}},
  \emph{\bibinfo{title}{The order parameter for spin glasses: a function on the
  interval $0$--$1$}}, \bibinfo{journal}{J. Phys. A}
  \textbf{\bibinfo{volume}{13}}, \bibinfo{pages}{1101} (\bibinfo{year}{1980}).

\bibitem[{\citenamefont{Parisi}(1983)}]{parisi:83}
\bibinfo{author}{\bibfnamefont{G.}~\bibnamefont{Parisi}},
  \emph{\bibinfo{title}{Order parameter for spin-glasses}},
  \bibinfo{journal}{Phys. Rev. Lett.} \textbf{\bibinfo{volume}{50}},
  \bibinfo{pages}{1946} (\bibinfo{year}{1983}).

\bibitem[{\citenamefont{Rammal et~al.}(1986)\citenamefont{Rammal, Toulouse, and
  Virasoro}}]{rammal:86}
\bibinfo{author}{\bibfnamefont{R.}~\bibnamefont{Rammal}},
  \bibinfo{author}{\bibfnamefont{G.}~\bibnamefont{Toulouse}}, \bibnamefont{and}
  \bibinfo{author}{\bibfnamefont{M.~A.} \bibnamefont{Virasoro}},
  \emph{\bibinfo{title}{{Ultrametricity for physicists}}},
  \bibinfo{journal}{Rev. Mod. Phys.} \textbf{\bibinfo{volume}{58}},
  \bibinfo{pages}{765} (\bibinfo{year}{1986}).

\bibitem[{\citenamefont{M\'ezard et~al.}(1987)\citenamefont{M\'ezard, Parisi,
  and Virasoro}}]{mezard:87}
\bibinfo{author}{\bibfnamefont{M.}~\bibnamefont{M\'ezard}},
  \bibinfo{author}{\bibfnamefont{G.}~\bibnamefont{Parisi}}, \bibnamefont{and}
  \bibinfo{author}{\bibfnamefont{M.~A.} \bibnamefont{Virasoro}},
  \emph{\bibinfo{title}{Spin Glass Theory and Beyond}}
  (\bibinfo{publisher}{World Scientific}, \bibinfo{address}{Singapore},
  \bibinfo{year}{1987}).

\bibitem[{\citenamefont{Parisi}(2008)}]{parisi:08}
\bibinfo{author}{\bibfnamefont{G.}~\bibnamefont{Parisi}},
  \emph{\bibinfo{title}{{{Some considerations of finite dimensional spin
  glasses}}}}, \bibinfo{journal}{J. Phys. A} \textbf{\bibinfo{volume}{41}},
  \bibinfo{pages}{324002} (\bibinfo{year}{2008}).

\bibitem[{\citenamefont{Sherrington and Kirkpatrick}(1975)}]{sherrington:75}
\bibinfo{author}{\bibfnamefont{D.}~\bibnamefont{Sherrington}} \bibnamefont{and}
  \bibinfo{author}{\bibfnamefont{S.}~\bibnamefont{Kirkpatrick}},
  \emph{\bibinfo{title}{Solvable model of a spin glass}},
  \bibinfo{journal}{Phys. Rev. Lett.} \textbf{\bibinfo{volume}{35}},
  \bibinfo{pages}{1792} (\bibinfo{year}{1975}).

\bibitem[{\citenamefont{McMillan}(1984)}]{mcmillan:84}
\bibinfo{author}{\bibfnamefont{W.~L.} \bibnamefont{McMillan}},
  \emph{\bibinfo{title}{Domain-wall renormalization-group study of the
  three-dimensional random {I}sing model}}, \bibinfo{journal}{Phys. Rev. B}
  \textbf{\bibinfo{volume}{30}}, \bibinfo{pages}{R476} (\bibinfo{year}{1984}).

\bibitem[{\citenamefont{Bray and Moore}(1986)}]{bray:86}
\bibinfo{author}{\bibfnamefont{A.~J.} \bibnamefont{Bray}} \bibnamefont{and}
  \bibinfo{author}{\bibfnamefont{M.~A.} \bibnamefont{Moore}},
  \emph{\bibinfo{title}{Scaling theory of the ordered phase of spin glasses}},
  in \emph{\bibinfo{booktitle}{Heidelberg Colloquium on Glassy Dynamics and
  Optimization}}, edited by
  \bibinfo{editor}{\bibfnamefont{L.}~\bibnamefont{Van~Hemmen}}
  \bibnamefont{and}
  \bibinfo{editor}{\bibfnamefont{I.}~\bibnamefont{Morgenstern}}
  (\bibinfo{publisher}{Springer}, \bibinfo{address}{New York},
  \bibinfo{year}{1986}), p. \bibinfo{pages}{121}.

\bibitem[{\citenamefont{Fisher and Huse}(1986)}]{fisher:86}
\bibinfo{author}{\bibfnamefont{D.~S.} \bibnamefont{Fisher}} \bibnamefont{and}
  \bibinfo{author}{\bibfnamefont{D.~A.} \bibnamefont{Huse}},
  \emph{\bibinfo{title}{Ordered phase of short-range {I}sing spin-glasses}},
  \bibinfo{journal}{Phys. Rev. Lett.} \textbf{\bibinfo{volume}{56}},
  \bibinfo{pages}{1601} (\bibinfo{year}{1986}).

\bibitem[{\citenamefont{Fisher and Huse}(1987)}]{fisher:87}
\bibinfo{author}{\bibfnamefont{D.~S.} \bibnamefont{Fisher}} \bibnamefont{and}
  \bibinfo{author}{\bibfnamefont{D.~A.} \bibnamefont{Huse}},
  \emph{\bibinfo{title}{Absence of many states in realistic spin glasses}},
  \bibinfo{journal}{J. Phys. A} \textbf{\bibinfo{volume}{20}},
  \bibinfo{pages}{L1005} (\bibinfo{year}{1987}).

\bibitem[{\citenamefont{Fisher and Huse}(1988)}]{fisher:88}
\bibinfo{author}{\bibfnamefont{D.~S.} \bibnamefont{Fisher}} \bibnamefont{and}
  \bibinfo{author}{\bibfnamefont{D.~A.} \bibnamefont{Huse}},
  \emph{\bibinfo{title}{Equilibrium behavior of the spin-glass ordered phase}},
  \bibinfo{journal}{Phys. Rev. B} \textbf{\bibinfo{volume}{38}},
  \bibinfo{pages}{386} (\bibinfo{year}{1988}).

\bibitem[{\citenamefont{de~Almeida and Thouless}(1978)}]{almeida:78}
\bibinfo{author}{\bibfnamefont{J.~R.~L.} \bibnamefont{de~Almeida}}
  \bibnamefont{and} \bibinfo{author}{\bibfnamefont{D.~J.}
  \bibnamefont{Thouless}}, \emph{\bibinfo{title}{Stability of the
  {S}herrington-{K}irkpatrick solution of a spin glass model}},
  \bibinfo{journal}{J. Phys. A} \textbf{\bibinfo{volume}{11}},
  \bibinfo{pages}{983} (\bibinfo{year}{1978}).

\bibitem[{\citenamefont{{Moore} and {Bray}}(2011)}]{moore:11}
\bibinfo{author}{\bibfnamefont{M.~A.} \bibnamefont{{Moore}}} \bibnamefont{and}
  \bibinfo{author}{\bibfnamefont{A.~J.} \bibnamefont{{Bray}}},
  \emph{\bibinfo{title}{{{Disappearance of the de Almeida-Thouless line in six
  dimensions}}}}, \bibinfo{journal}{Phys. Rev. B}
  \textbf{\bibinfo{volume}{83}}, \bibinfo{pages}{224408}
  (\bibinfo{year}{2011}).

\bibitem[{\citenamefont{{Moore}}(2012)}]{moore:12}
\bibinfo{author}{\bibfnamefont{M.~A.} \bibnamefont{{Moore}}},
  \emph{\bibinfo{title}{{{{$1/m$} expansion in spin glasses and the de
  Almeida-Thouless line}}}}, \bibinfo{journal}{Phys. Rev. E}
  \textbf{\bibinfo{volume}{86}}, \bibinfo{pages}{031114}
  (\bibinfo{year}{2012}).

\bibitem[{\citenamefont{Bray and Roberts}(1980)}]{bray:80b}
\bibinfo{author}{\bibfnamefont{A.~J.} \bibnamefont{Bray}} \bibnamefont{and}
  \bibinfo{author}{\bibfnamefont{S.~A.} \bibnamefont{Roberts}},
  \emph{\bibinfo{title}{Renormalisation-group approach to the spin glass
  transition in finite magnetic field}}, \bibinfo{journal}{J. Phys. C}
  \textbf{\bibinfo{volume}{13}}, \bibinfo{pages}{5405} (\bibinfo{year}{1980}).

\bibitem[{\citenamefont{{Kotliar} et~al.}(1983)\citenamefont{{Kotliar},
  {Anderson}, and {Stein}}}]{kotliar:83}
\bibinfo{author}{\bibfnamefont{G.}~\bibnamefont{{Kotliar}}},
  \bibinfo{author}{\bibfnamefont{P.~W.} \bibnamefont{{Anderson}}},
  \bibnamefont{and} \bibinfo{author}{\bibfnamefont{D.~L.}
  \bibnamefont{{Stein}}}, \emph{\bibinfo{title}{One-dimensional spin-glass
  model with long-range random interactions}}, \bibinfo{journal}{Phys. Rev. B}
  \textbf{\bibinfo{volume}{27}}, \bibinfo{pages}{602} (\bibinfo{year}{1983}).

\bibitem[{\citenamefont{Katzgraber and
  Young}(2003{\natexlab{a}})}]{katzgraber:03}
\bibinfo{author}{\bibfnamefont{H.~G.} \bibnamefont{Katzgraber}}
  \bibnamefont{and} \bibinfo{author}{\bibfnamefont{A.~P.} \bibnamefont{Young}},
  \emph{\bibinfo{title}{Monte {C}arlo studies of the one-dimensional {I}sing
  spin glass with power-law interactions}}, \bibinfo{journal}{Phys. Rev. B}
  \textbf{\bibinfo{volume}{67}}, \bibinfo{pages}{134410}
  (\bibinfo{year}{2003}{\natexlab{a}}).

\bibitem[{\citenamefont{Bray et~al.}(1986)\citenamefont{Bray, Moore, and
  Young}}]{bray:86b}
\bibinfo{author}{\bibfnamefont{A.~J.} \bibnamefont{Bray}},
  \bibinfo{author}{\bibfnamefont{M.~A.} \bibnamefont{Moore}}, \bibnamefont{and}
  \bibinfo{author}{\bibfnamefont{A.~P.} \bibnamefont{Young}},
  \emph{\bibinfo{title}{Lower critical dimension of metallic vector
  spin-glasses}}, \bibinfo{journal}{Phys. Rev. Lett}
  \textbf{\bibinfo{volume}{56}}, \bibinfo{pages}{2641} (\bibinfo{year}{1986}).

\bibitem[{\citenamefont{Katzgraber et~al.}(2009)\citenamefont{Katzgraber,
  Larson, and Young}}]{katzgraber:09b}
\bibinfo{author}{\bibfnamefont{H.~G.} \bibnamefont{Katzgraber}},
  \bibinfo{author}{\bibfnamefont{D.}~\bibnamefont{Larson}}, \bibnamefont{and}
  \bibinfo{author}{\bibfnamefont{A.~P.} \bibnamefont{Young}},
  \emph{\bibinfo{title}{Study of the de {A}lmeida-{T}houless line using
  power-law diluted one-dimensional {I}sing spin glasses}},
  \bibinfo{journal}{Phys. Rev. Lett.} \textbf{\bibinfo{volume}{102}},
  \bibinfo{pages}{177205} (\bibinfo{year}{2009}).

\bibitem[{\citenamefont{{Leuzzi} et~al.}(2009)\citenamefont{{Leuzzi}, {Parisi},
  {Ricci-Tersenghi}, and {Ruiz-Lorenzo}}}]{leuzzi:09}
\bibinfo{author}{\bibfnamefont{L.}~\bibnamefont{{Leuzzi}}},
  \bibinfo{author}{\bibfnamefont{G.}~\bibnamefont{{Parisi}}},
  \bibinfo{author}{\bibfnamefont{F.}~\bibnamefont{{Ricci-Tersenghi}}},
  \bibnamefont{and} \bibinfo{author}{\bibfnamefont{J.~J.}
  \bibnamefont{{Ruiz-Lorenzo}}}, \emph{\bibinfo{title}{{{Ising Spin-Glass
  Transition in a Magnetic Field Outside the Limit of Validity of Mean-Field
  Theory}}}}, \bibinfo{journal}{Phys. Rev. Lett.}
  \textbf{\bibinfo{volume}{103}}, \bibinfo{pages}{267201}
  (\bibinfo{year}{2009}).

\bibitem[{\citenamefont{{Ba{\~n}os} et~al.}(2012)\citenamefont{{Ba{\~n}os},
  {Fernandez}, {Martin-Mayor}, and {Young}}}]{banos:12b}
\bibinfo{author}{\bibfnamefont{R.~A.} \bibnamefont{{Ba{\~n}os}}},
  \bibinfo{author}{\bibfnamefont{L.~A.} \bibnamefont{{Fernandez}}},
  \bibinfo{author}{\bibfnamefont{V.}~\bibnamefont{{Martin-Mayor}}},
  \bibnamefont{and} \bibinfo{author}{\bibfnamefont{A.~P.}
  \bibnamefont{{Young}}}, \emph{\bibinfo{title}{{{Correspondence between
  long-range and short-range spin glasses}}}}, \bibinfo{journal}{Phys. Rev. B}
  \textbf{\bibinfo{volume}{86}}, \bibinfo{pages}{134416}
  (\bibinfo{year}{2012}).

\bibitem[{\citenamefont{Aspelmeier et~al.}(2016)\citenamefont{Aspelmeier,
  Katzgraber, Larson, Moore, Wittmann, and Yeo}}]{aspelmeier:16}
\bibinfo{author}{\bibfnamefont{T.}~\bibnamefont{Aspelmeier}},
  \bibinfo{author}{\bibfnamefont{H.~G.} \bibnamefont{Katzgraber}},
  \bibinfo{author}{\bibfnamefont{D.}~\bibnamefont{Larson}},
  \bibinfo{author}{\bibfnamefont{M.~A.} \bibnamefont{Moore}},
  \bibinfo{author}{\bibfnamefont{M.}~\bibnamefont{Wittmann}}, \bibnamefont{and}
  \bibinfo{author}{\bibfnamefont{J.}~\bibnamefont{Yeo}},
  \emph{\bibinfo{title}{{Finite-size critical scaling in Ising spin glasses in
  the mean-field regime}}}, \bibinfo{journal}{Phys. Rev. E}
  \textbf{\bibinfo{volume}{93}}, \bibinfo{pages}{032123}
  (\bibinfo{year}{2016}).

\bibitem[{\citenamefont{Moore}(2010)}]{moore:10}
\bibinfo{author}{\bibfnamefont{M.~A.} \bibnamefont{Moore}},
  \emph{\bibinfo{title}{{{Ordered phase of the one-dimensional Ising spin glass
  with long-range interactions}}}}, \bibinfo{journal}{Phys. Rev. B}
  \textbf{\bibinfo{volume}{82}}, \bibinfo{pages}{014417}
  (\bibinfo{year}{2010}).

\bibitem[{\citenamefont{Southern and Young}(1977)}]{southern:77}
\bibinfo{author}{\bibfnamefont{B.~W.} \bibnamefont{Southern}} \bibnamefont{and}
  \bibinfo{author}{\bibfnamefont{A.~P.} \bibnamefont{Young}},
  \emph{\bibinfo{title}{{Real space rescaling study of spin glass behaviour in
  three dimensions}}}, \bibinfo{journal}{J. Phys. C}
  \textbf{\bibinfo{volume}{10}}, \bibinfo{pages}{2179} (\bibinfo{year}{1977}).

\bibitem[{\citenamefont{Bray and Moore}(1984)}]{bray:84}
\bibinfo{author}{\bibfnamefont{A.~J.} \bibnamefont{Bray}} \bibnamefont{and}
  \bibinfo{author}{\bibfnamefont{M.~A.} \bibnamefont{Moore}},
  \emph{\bibinfo{title}{Lower critical dimension of {I}sing spin glasses: a
  numerical study}}, \bibinfo{journal}{J. Phys. C}
  \textbf{\bibinfo{volume}{17}}, \bibinfo{pages}{L463} (\bibinfo{year}{1984}).

\bibitem[{\citenamefont{Parisi and Rizzo}(2010)}]{parisi:10}
\bibinfo{author}{\bibfnamefont{G.}~\bibnamefont{Parisi}} \bibnamefont{and}
  \bibinfo{author}{\bibfnamefont{T.}~\bibnamefont{Rizzo}},
  \emph{\bibinfo{title}{{Universality and deviations in disordered systems}}},
  \bibinfo{journal}{Phys. Rev. B} \textbf{\bibinfo{volume}{81}},
  \bibinfo{pages}{094201} (\bibinfo{year}{2010}).

\bibitem[{\citenamefont{Wehr and Aizenman}(1990)}]{wehr:90}
\bibinfo{author}{\bibfnamefont{J.}~\bibnamefont{Wehr}} \bibnamefont{and}
  \bibinfo{author}{\bibfnamefont{M.}~\bibnamefont{Aizenman}},
  \emph{\bibinfo{title}{{Fluctuations of Extensive Functions of Quenched Random
  Couplings}}}, \bibinfo{journal}{J. Stat. Phys.}
  \textbf{\bibinfo{volume}{60}}, \bibinfo{pages}{287} (\bibinfo{year}{1990}).

\bibitem[{\citenamefont{Banavar and Cieplak}(1982)}]{banavar:82}
\bibinfo{author}{\bibfnamefont{J.~R.} \bibnamefont{Banavar}} \bibnamefont{and}
  \bibinfo{author}{\bibfnamefont{M.}~\bibnamefont{Cieplak}},
  \emph{\bibinfo{title}{{Nature of Ordering in Spin-Glasses}}},
  \bibinfo{journal}{Phys. Rev. Lett.} \textbf{\bibinfo{volume}{48}},
  \bibinfo{pages}{832} (\bibinfo{year}{1982}).

\bibitem[{\citenamefont{Aspelmeier et~al.}(2003)\citenamefont{Aspelmeier,
  Moore, and Young}}]{aspelmeier:02}
\bibinfo{author}{\bibfnamefont{T.}~\bibnamefont{Aspelmeier}},
  \bibinfo{author}{\bibfnamefont{M.~A.} \bibnamefont{Moore}}, \bibnamefont{and}
  \bibinfo{author}{\bibfnamefont{A.~P.} \bibnamefont{Young}},
  \emph{\bibinfo{title}{Interface energies in {I}sing spin glasses}},
  \bibinfo{journal}{Phys. Rev. Lett.} \textbf{\bibinfo{volume}{90}},
  \bibinfo{pages}{127202} (\bibinfo{year}{2003}).

\bibitem[{\citenamefont{de~Dominicis et~al.}(1998)\citenamefont{de~Dominicis,
  Kondor, and Temesv{\'a}ri}}]{dedominicis:98}
\bibinfo{author}{\bibfnamefont{C.}~\bibnamefont{de~Dominicis}},
  \bibinfo{author}{\bibfnamefont{I.}~\bibnamefont{Kondor}}, \bibnamefont{and}
  \bibinfo{author}{\bibfnamefont{T.}~\bibnamefont{Temesv{\'a}ri}},
  \emph{\bibinfo{title}{Beyond the sherrington-kirkpatrick model}}, in
  \emph{\bibinfo{booktitle}{Spin Glasses and Random Fields}}, edited by
  \bibinfo{editor}{\bibfnamefont{A.}~\bibnamefont{Young}}
  (\bibinfo{publisher}{World Scientific}, \bibinfo{address}{Singapore},
  \bibinfo{year}{1998}).

\bibitem[{\citenamefont{Katzgraber and {Young}}(2005)}]{katzgraber:05c}
\bibinfo{author}{\bibfnamefont{H.~G.} \bibnamefont{Katzgraber}}
  \bibnamefont{and} \bibinfo{author}{\bibfnamefont{A.~P.}
  \bibnamefont{{Young}}}, \emph{\bibinfo{title}{{{Probing the Almeida-Thouless
  line away from the mean-field model}}}}, \bibinfo{journal}{Phys. Rev. B}
  \textbf{\bibinfo{volume}{72}}, \bibinfo{pages}{184416}
  (\bibinfo{year}{2005}).

\bibitem[{\citenamefont{Marinari and Parisi}(2000)}]{marinari:00}
\bibinfo{author}{\bibfnamefont{E.}~\bibnamefont{Marinari}} \bibnamefont{and}
  \bibinfo{author}{\bibfnamefont{G.}~\bibnamefont{Parisi}},
  \emph{\bibinfo{title}{On the effects of changing the boundary conditions on
  the ground state of {I}sing spin glasses}}, \bibinfo{journal}{Phys. Rev. B}
  \textbf{\bibinfo{volume}{62}}, \bibinfo{pages}{11677} (\bibinfo{year}{2000}).

\bibitem[{\citenamefont{Aspelmeier and Moore}(2003)}]{aspelmeier:03}
\bibinfo{author}{\bibfnamefont{T.}~\bibnamefont{Aspelmeier}} \bibnamefont{and}
  \bibinfo{author}{\bibfnamefont{M.~A.} \bibnamefont{Moore}},
  \emph{\bibinfo{title}{{Free Energy Fluctuations in Ising Spin Glasses}}},
  \bibinfo{journal}{Phys. Rev. Lett.} \textbf{\bibinfo{volume}{90}},
  \bibinfo{pages}{177201} (\bibinfo{year}{2003}).

\bibitem[{\citenamefont{Aspelmeier}(2008{\natexlab{a}})}]{aspelmeier:08a}
\bibinfo{author}{\bibfnamefont{T.}~\bibnamefont{Aspelmeier}},
  \emph{\bibinfo{title}{{Free-Energy Fluctuations and Chaos in the
  Sherrington-Kirkpatrick Model}}}, \bibinfo{journal}{Phys. Rev. Lett.}
  \textbf{\bibinfo{volume}{100}}, \bibinfo{pages}{117205}
  (\bibinfo{year}{2008}{\natexlab{a}}).

\bibitem[{\citenamefont{Aspelmeier}(2008{\natexlab{b}})}]{aspelmeier:08b}
\bibinfo{author}{\bibfnamefont{T.}~\bibnamefont{Aspelmeier}},
  \emph{\bibinfo{title}{{An exact relation between free energy fluctuations and
  bond chaos in the Sherrington-Kirkpatrick model}}}, \bibinfo{journal}{J.
  Stat. Mech.} \textbf{\bibinfo{volume}{\normalfont{P04018}}}
  (\bibinfo{year}{2008}{\natexlab{b}}).

\bibitem[{\citenamefont{Aspelmeier et~al.}(2008)\citenamefont{Aspelmeier,
  Billoire, Marinari, and Moore}}]{aspelmeier:08}
\bibinfo{author}{\bibfnamefont{T.}~\bibnamefont{Aspelmeier}},
  \bibinfo{author}{\bibfnamefont{A.}~\bibnamefont{Billoire}},
  \bibinfo{author}{\bibfnamefont{E.}~\bibnamefont{Marinari}}, \bibnamefont{and}
  \bibinfo{author}{\bibfnamefont{M.~A.} \bibnamefont{Moore}},
  \emph{\bibinfo{title}{{{Finite-size corrections in the Sherrington
  Kirkpatrick model}}}}, \bibinfo{journal}{J. Phys. A: Math. Theor.}
  \textbf{\bibinfo{volume}{41}}, \bibinfo{pages}{324008}
  (\bibinfo{year}{2008}).

\bibitem[{\citenamefont{Wang et~al.}(2016)\citenamefont{Wang, Katzgraber, and
  Machta}}]{wang:16x}
\bibinfo{author}{\bibfnamefont{W.}~\bibnamefont{Wang}},
  \bibinfo{author}{\bibfnamefont{H.~G.} \bibnamefont{Katzgraber}},
  \bibnamefont{and} \bibinfo{author}{\bibfnamefont{J.}~\bibnamefont{Machta}},
  \emph{\bibinfo{title}{{On the number of thermodynamics states in the
  Edwards-Anderson Ising spin glass}}}, \bibinfo{journal}{in preparation}
  (\bibinfo{year}{2016}).

\bibitem[{\citenamefont{{Moore}}(2005)}]{moore:05}
\bibinfo{author}{\bibfnamefont{M.~A.} \bibnamefont{{Moore}}},
  \emph{\bibinfo{title}{{{The stability of the replica-symmetric state in
  finite-dimensional spin glasses}}}}, \bibinfo{journal}{J. Phys. A}
  \textbf{\bibinfo{volume}{38}}, \bibinfo{pages}{L783} (\bibinfo{year}{2005}).

\bibitem[{\citenamefont{Boettcher}(2005{\natexlab{a}})}]{boettcher:05d}
\bibinfo{author}{\bibfnamefont{S.}~\bibnamefont{Boettcher}},
  \emph{\bibinfo{title}{{Stiffness of the {E}dwards-{A}nderson Model in all
  Dimensions}}}, \bibinfo{journal}{Phys. Rev. Lett.}
  \textbf{\bibinfo{volume}{95}}, \bibinfo{pages}{197205}
  (\bibinfo{year}{2005}{\natexlab{a}}).

\bibitem[{\citenamefont{Boettcher}(2004)}]{boettcher:04b}
\bibinfo{author}{\bibfnamefont{S.}~\bibnamefont{Boettcher}},
  \emph{\bibinfo{title}{{Low-temperature excitations of dilute lattice spin
  glasses}}}, \bibinfo{journal}{Europhys. Lett.} \textbf{\bibinfo{volume}{67}},
  \bibinfo{pages}{453} (\bibinfo{year}{2004}).

\bibitem[{\citenamefont{Katzgraber and
  Young}(2003{\natexlab{b}})}]{katzgraber:03f}
\bibinfo{author}{\bibfnamefont{H.~G.} \bibnamefont{Katzgraber}}
  \bibnamefont{and} \bibinfo{author}{\bibfnamefont{A.~P.} \bibnamefont{Young}},
  \emph{\bibinfo{title}{Geometry of large-scale low-energy excitations in the
  one-dimensional {I}sing spin glass with power-law interactions}},
  \bibinfo{journal}{Phys. Rev. B} \textbf{\bibinfo{volume}{68}},
  \bibinfo{pages}{224408} (\bibinfo{year}{2003}{\natexlab{b}}).

\bibitem[{\citenamefont{Boettcher}(2005{\natexlab{b}})}]{boettcher:05}
\bibinfo{author}{\bibfnamefont{S.}~\bibnamefont{Boettcher}},
  \emph{\bibinfo{title}{{Extremal Optimization for Sherrington-Kirkpatrick Spin
  Glasses}}}, \bibinfo{journal}{E. Phys. J. B} \textbf{\bibinfo{volume}{46}},
  \bibinfo{pages}{501} (\bibinfo{year}{2005}{\natexlab{b}}).

\bibitem[{\citenamefont{Katzgraber et~al.}(2005)\citenamefont{Katzgraber,
  {K{\"o}rner}, {Liers}, and {Hartmann}}}]{katzgraber:05d}
\bibinfo{author}{\bibfnamefont{H.~G.} \bibnamefont{Katzgraber}},
  \bibinfo{author}{\bibfnamefont{M.}~\bibnamefont{{K{\"o}rner}}},
  \bibinfo{author}{\bibfnamefont{F.}~\bibnamefont{{Liers}}}, \bibnamefont{and}
  \bibinfo{author}{\bibfnamefont{A.~K.} \bibnamefont{{Hartmann}}},
  \emph{\bibinfo{title}{{Overcoming System-Size Limitations in Spin Glasses}}},
  \bibinfo{journal}{Prog. Theor. Phys. Suppl.} \textbf{\bibinfo{volume}{157}},
  \bibinfo{pages}{59} (\bibinfo{year}{2005}).

\bibitem[{\citenamefont{Boettcher}(2010)}]{boettcher:10}
\bibinfo{author}{\bibfnamefont{S.}~\bibnamefont{Boettcher}},
  \emph{\bibinfo{title}{{Simulations of ground state fluctuations in mean-field
  Ising spin glasses}}}, \bibinfo{journal}{J. Stat. Mech.}
  \textbf{\bibinfo{volume}{\normalfont{P07002}}} (\bibinfo{year}{2010}).

\bibitem[{\citenamefont{{Belletti} et~al.}(2008)\citenamefont{{Belletti},
  {Cotallo}, {Cruz}, {Fern{\'a}ndez}, {Gordillo}, {Maiorano}, {Mantovani},
  {Marinari}, {Mart{\'{\i}}n-Mayor}, {Mu{\~n}oz-Sudupe} et~al.}}]{belletti:08}
\bibinfo{author}{\bibfnamefont{F.}~\bibnamefont{{Belletti}}},
  \bibinfo{author}{\bibfnamefont{M.}~\bibnamefont{{Cotallo}}},
  \bibinfo{author}{\bibfnamefont{A.}~\bibnamefont{{Cruz}}},
  \bibinfo{author}{\bibfnamefont{L.~A.} \bibnamefont{{Fern{\'a}ndez}}},
  \bibinfo{author}{\bibfnamefont{A.}~\bibnamefont{{Gordillo}}},
  \bibinfo{author}{\bibfnamefont{A.}~\bibnamefont{{Maiorano}}},
  \bibinfo{author}{\bibfnamefont{F.}~\bibnamefont{{Mantovani}}},
  \bibinfo{author}{\bibfnamefont{E.}~\bibnamefont{{Marinari}}},
  \bibinfo{author}{\bibfnamefont{V.}~\bibnamefont{{Mart{\'{\i}}n-Mayor}}},
  \bibinfo{author}{\bibfnamefont{A.}~\bibnamefont{{Mu{\~n}oz-Sudupe}}},
  \bibnamefont{et~al.}, \emph{\bibinfo{title}{{{Simulating spin systems on
  IANUS, an FPGA-based computer}}}}, \bibinfo{journal}{Comp. Phys. Comm.}
  \textbf{\bibinfo{volume}{178}}, \bibinfo{pages}{208} (\bibinfo{year}{2008}).

\bibitem[{\citenamefont{{Leuzzi} et~al.}(2008)\citenamefont{{Leuzzi}, {Parisi},
  {Ricci-Tersenghi}, and {Ruiz-Lorenzo}}}]{leuzzi:08}
\bibinfo{author}{\bibfnamefont{L.}~\bibnamefont{{Leuzzi}}},
  \bibinfo{author}{\bibfnamefont{G.}~\bibnamefont{{Parisi}}},
  \bibinfo{author}{\bibfnamefont{F.}~\bibnamefont{{Ricci-Tersenghi}}},
  \bibnamefont{and} \bibinfo{author}{\bibfnamefont{J.~J.}
  \bibnamefont{{Ruiz-Lorenzo}}}, \emph{\bibinfo{title}{{{Diluted
  One-Dimensional Spin Glasses with Power Law Decaying Interactions}}}},
  \bibinfo{journal}{Phys. Rev. Lett.} \textbf{\bibinfo{volume}{101}},
  \bibinfo{pages}{107203} (\bibinfo{year}{2008}).

\bibitem[{\citenamefont{Hukushima and Iba}(2003)}]{hukushima:03}
\bibinfo{author}{\bibfnamefont{K.}~\bibnamefont{Hukushima}} \bibnamefont{and}
  \bibinfo{author}{\bibfnamefont{Y.}~\bibnamefont{Iba}}, in
  \emph{\bibinfo{booktitle}{{The Monte Carlo method in the physical sciences:
  celebrating the 50th anniversary of the Metropolis algorithm}}}, edited by
  \bibinfo{editor}{\bibfnamefont{J.~E.} \bibnamefont{Gubernatis}}
  (\bibinfo{publisher}{AIP}, \bibinfo{year}{2003}), vol. \bibinfo{volume}{690},
  p. \bibinfo{pages}{200}.

\bibitem[{\citenamefont{Zhou and Chen}(2010)}]{zhou:10}
\bibinfo{author}{\bibfnamefont{E.}~\bibnamefont{Zhou}} \bibnamefont{and}
  \bibinfo{author}{\bibfnamefont{X.}~\bibnamefont{Chen}}, in
  \emph{\bibinfo{booktitle}{{Proceedings of the 2010 Winter Simulation
  Conference (WSC)}}} (\bibinfo{publisher}{Springer},
  \bibinfo{address}{Baltimore MD}, \bibinfo{year}{2010}), p.
  \bibinfo{pages}{1211}.

\bibitem[{\citenamefont{Machta}(2010)}]{machta:10}
\bibinfo{author}{\bibfnamefont{J.}~\bibnamefont{Machta}},
  \emph{\bibinfo{title}{{Population annealing with weighted averages: A {M}onte
  {C}arlo method for rough free-energy landscapes}}}, \bibinfo{journal}{Phys.
  Rev. E} \textbf{\bibinfo{volume}{82}}, \bibinfo{pages}{026704}
  (\bibinfo{year}{2010}).

\bibitem[{\citenamefont{Wang et~al.}(2015)\citenamefont{Wang, Machta, and
  Katzgraber}}]{wang:15e}
\bibinfo{author}{\bibfnamefont{W.}~\bibnamefont{Wang}},
  \bibinfo{author}{\bibfnamefont{J.}~\bibnamefont{Machta}}, \bibnamefont{and}
  \bibinfo{author}{\bibfnamefont{H.~G.} \bibnamefont{Katzgraber}},
  \emph{\bibinfo{title}{{Population annealing: Theory and application in spin
  glasses}}}, \bibinfo{journal}{Phys. Rev. E} \textbf{\bibinfo{volume}{92}},
  \bibinfo{pages}{063307} (\bibinfo{year}{2015}).

\end{thebibliography}

\end{document}